\documentclass[sigconf]{acmart}
\AtBeginDocument{%
  }

\copyrightyear{2026}
\acmYear{2026}
\setcopyright{cc}
\setcctype{by}
\acmConference[DIS '26]{Designing Interactive Systems Conference}{June 13--17, 2026}{Singapore, Singapore}
\acmBooktitle{Designing Interactive Systems Conference (DIS '26), June 13--17, 2026, Singapore, Singapore}
\acmDOI{10.1145/3800645.3813102}
\acmISBN{979-8-4007-2563-0/2026/06}

\begin{document}

%%
%% The ``title'' command has an optional parameter,
%% allowing the author to define a ``short title'' to be used in page headers.
\title{Guidelines for Designing AI Technologies to Support Adult Learning}

%%
%% The ``author'' command and its associated commands are used to define
%% the authors and their affiliations.
%% Of note is the shared affiliation of the first two authors, and the
%% ``authornote'' and ``authornotemark'' commands
%% used to denote shared contribution to the research.
\author{Jennifer M. Reddig}
\authornote{Both authors contributed equally to this research.}
\affiliation{%
  \institution{Georgia Institute of Technology}
  \city{Atlanta}
  \state{Georgia}
  \country{USA}
}

\author{Glen R. Smith Jr.}
\authornotemark[1]
\affiliation{%
  \institution{Georgia Institute of Technology}
  \city{Atlanta}
  \state{Georgia}
  \country{USA}
}

\author{Sanaz Ahmadzadeh Siyahrood}
\affiliation{%
  \institution{Georgia Institute of Technology}
  \city{Atlanta}
  \state{Georgia}
  \country{USA}
}

\author{Wesley G. Morris}
\affiliation{%
  \institution{Vanderbilt University}
  \city{Nashville}
  \state{Tennessee}
  \country{USA}
}

\author{Yoojin Bae}
\affiliation{%
  \institution{Georgia State University}
  \city{Atlanta}
  \state{Georgia}
  \country{USA}
}

\author{Kaitlyn Crutcher}
\affiliation{%
  \institution{Georgia Institute of Technology}
  \city{Atlanta}
  \state{Georgia}
  \country{USA}
}

\author{John Kos}
\affiliation{%
  \institution{Georgia Institute of Technology}
  \city{Atlanta}
  \state{Georgia}
  \country{USA}
}

\author{Rahul K. Dass}
\affiliation{%
  \institution{Georgia Institute of Technology}
  \city{Atlanta}
  \state{Georgia}
  \country{USA}
}

\author{Jinho Kim}
\affiliation{%
  \institution{Georgia State University}
  \city{Atlanta}
  \state{Georgia}
  \country{USA}
}

\author{Momin Naushad Siddiqui}
\affiliation{%
  \institution{Georgia Institute of Technology}
  \city{Atlanta}
  \state{Georgia}
  \country{USA}
}

\author{Daniel Weitekamp}
\affiliation{%
  \institution{Georgia Institute of Technology}
  \city{Atlanta}
  \state{Georgia}
  \country{USA}
}

\author{Ploy Thajchayapong}
\affiliation{%
  \institution{Georgia Institute of Technology}
  \city{Atlanta}
  \state{Georgia}
  \country{USA}
}

\author{Sandeep Kakar}
\affiliation{%
  \institution{Georgia Institute of Technology}
  \city{Atlanta}
  \state{Georgia}
  \country{USA}
}

\author{Alex Endert}
\affiliation{%
  \institution{Georgia Institute of Technology}
  \city{Atlanta}
  \state{Georgia}
  \country{USA}
}

\author{Scott Crossley}
\affiliation{%
  \institution{Vanderbilt University}
  \city{Nashville}
  \state{Tennessee}
  \country{USA}
}

\author{Min Kyu Kim}
\affiliation{%
  \institution{Georgia State University}
  \city{Atlanta}
  \state{Georgia}
  \country{USA}
}

\author{Chris Dede}
\affiliation{%
  \institution{Harvard University}
  \city{Cambridge}
  \state{Massachusetts}
  \country{USA}
}

\author{Ashok Goel}
\affiliation{%
  \institution{Georgia Institute of Technology}
  \city{Atlanta}
  \state{Georgia}
  \country{USA}
}

\author{Christopher J. MacLellan}
\affiliation{%
  \institution{Georgia Institute of Technology}
  \city{Atlanta}
  \state{Georgia}
  \country{USA}
}

%%
%% By default, the full list of authors will be used in the page
%% headers. Often, this list is too long, and will overlap
%% other information printed in the page headers. This command allows
%% the author to define a more concise list
%% of authors' names for this purpose.
\renewcommand{\shortauthors}{Reddig and Smith et al.}

%%
%% The abstract is a short summary of the work to be presented in the
%% article.
\begin{abstract}
AI-powered educational technologies have demonstrated measurable benefits for learners, but their design and evaluation have largely centered on K-12 contexts. As a result, many AI-supported learning systems remain poorly aligned with the needs, constraints, and goals of adult learners.
To better understand how AI systems function in adult education, 
this paper examines the deployment of several AI learning technologies developed within a multidisciplinary, national research institute in the United States focused on adult learning and online education. Drawing on longitudinal deployment data,
% including reports, focus group transcripts, design documentation, and feedback from learners, instructors, administrators, and research teams, 
we conducted a reflexive thematic analysis to identify recurring challenges and design considerations across systems. These insights were synthesized into a set of 19 design guidelines intended to inform future AI-supported adult learning technologies. We demonstrate the utility of these guidelines through a heuristic evaluation of the deployed systems.
% , illustrating both how existing tools can be assessed and how others might apply the process in their own work. 
Lastly, we present a guideline exploration tool that aids in the ideation of technologies by connecting the guidelines to stakeholder statements surfaced in the analysis process.
\end{abstract}

%%
%% The code below is generated by the tool at http://dl.acm.org/ccs.cfm.
%% Please copy and paste the code instead of the example below.
%%


%%
%% Keywords. The author(s) should pick words that accurately describe
%% the work being presented. Separate the keywords with commas.
%% A ``teaser'' image appears between the author and affiliation
%% information and the body of the document, and typically spans the
%% page.
% \begin{teaserfigure}
%   \includegraphics[width=\textwidth]{sampleteaser}
%   \caption{Seattle Mariners at Spring Training, 2010.}
%   \Description{Enjoying the baseball game from the third-base
%   seats. Ichiro Suzuki preparing to bat.}
%   \label{fig:teaser}
% \end{teaserfigure}

\received{19 January 2026}
\received[revised]{12 March 2026}
\received[accepted]{5 June 2026}

%%
%% This command processes the author and affiliation and title
%% information and builds the first part of the formatted document.
\maketitle

\section{Introduction}

AI-powered educational technologies enable adaptive, data-driven instruction across a wide range of learning contexts and have demonstrated measurable benefits for learner engagement, personalization, and instructional efficiency at scale \cite{chiu2023systematic}. Although recent trends in educational technology research reflect a growing emphasis on higher education and adult learning populations \cite{bond2019revisiting, kimmons2022trends}, many AI-supported systems continue to be designed and evaluated within K-12 settings, where learning typically occurs in formal classrooms with relatively stable schedules, curricula, and support structures. A systematic review by \citet{bernacki2021systematic} suggests that research on personalized learning has largely overlooked adult learning and workplace training populations. As a result, prevailing design knowledge, assumptions, and system evaluation criteria are tightly coupled to the needs and constraints of younger learners and traditional schooling contexts \cite{dube2022identification}. This renders many AI-supported learning technologies poorly aligned with the needs of adult learners, who increasingly pursue education through professional training and informal learning pathways \cite{meierkord2025trends}.

Adult learners differ from younger learners not only demographically, but motivationally and contextually, with learning situated within the broader constraints of their daily lives. Adult learning is often self-directed and goal-oriented, shaped by immediate needs such as career advancement, credentialing, or reskilling \cite{knowles1984andragogy, knowles1973adult}. At the same time, adult learners must balance education alongside employment, family responsibilities, and other social obligations, which constrains when, where, and how learning occurs. These conditions shape patterns of technology use and challenge design assumptions that presuppose continuous participation, uniform goals, or stable instructional environments. Consequently, learning technologies intended for adult learners must account for the contextual conditions under which learning takes place, rather than inheriting design patterns optimized for K-12 settings.

This paper examines these challenges through an empirical investigation of AI-powered learning technologies deployed in adult education contexts. Our analysis focuses on a set of systems developed and deployed within the National AI Institute for Adult Learning and Online Education (AI-ALOE) in the United States \cite{goel2024ai}. Even though these technologies were implemented and deployed across diverse adult learning environments, stakeholders like learners, instructors, administrators, and technology development teams repeatedly surfaced similar needs, challenges, and design considerations. The presence of these patterns across otherwise distinct deployments motivated a more systematic, cross-technology synthesis of findings.

Drawing on longitudinal deployment data---including focus groups with instructors and learners, technical artifacts, and progress reports---we analyze how these technologies were designed, implemented, and experienced in practice. Using reflexive thematic analysis \cite{braun2019reflecting}, we synthesize themes and insights from these deployments into a set of 19 design guidelines. We then demonstrate the applicability of these guidelines by evaluating them against the same set of technologies, highlighting opportunities for improvement and the trade-offs that emerge in real-world adult learning environments. Collectively, this work contributes:
\vspace{2pt} 

\begin{enumerate}
    \item An empirically grounded set of 19 guidelines for AI-powered adult learning technologies,
    \item A method for conducting heuristic evaluations of adult learning technologies using these guidelines, and
    \item A guideline exploration tool that informs meaningful ideation and technology refinement by linking guidelines to  stakeholder statements surfaced during our analysis. 
\end{enumerate}
\vspace{2pt}

We begin by situating this work within prior research on adult education and AI-supported learning technologies. We then describe the learning technologies examined in this study and the adult learning contexts in which they were deployed. Next, we detail our research methodology, including data sources and analytic procedures. We subsequently present the resulting design guidelines, followed by a discussion of their implications for the design and evaluation of learning technologies for adult learners and directions for future research.

%We propose that these design guidelines could be used for generating ideas for new technologies or problem-solving existing technology issues through a user-based lens.

% 3. AI-ALOE is uniquely positioned to solve this, AI technologies deployed in real-world classroom contexts (describe briefly the tasks that these technologies serve), ecosystem of teachers, students, COI, etc. probe adult learning through co-design and focus groups

% The NSF AI Institute for Adult Learning and Online Education (AI-ALOE) is one of twenty-nine AI Institutes supporting advancements in foundational AI.

% 4. bulleted list of our contributions
% - thematic analysis of various stakeholder perspectives across the different ALOE technologies
% - design guidelines developed from broad themes of user needs
% - heuristic evaluation of the guidelines, demonstrate applicability of guidelines by examining which are relevant/present in each tech
% propose how to use the guidelines to support lifelong learning technologies

\section{Related Work}

\citet{knowles1973adult} introduces andragogy, the theory of adult learning, which points out that adults are independent, self-directed learners with years of life experience who require meaningful, problem-oriented teaching strategies.
In contrast, Knowles describes pedagogy, the theory of child learning, as being characterized by a more dependent relationship between instructor and child, where the instructor largely shapes the direction and content of the curriculum. 
According to Knowles, adults require instructional strategies that are collaborative and demonstrate how content supports their learning goals.
Adults exist in a different social and cultural space than children, distinguished by a greater degree of self-directed decision-making. 
Educational technology for adults must support a different set of needs than tools designed for children \cite{tusting2003models}. 
%so education and educational tools need to support the unique adult learning context \cite{tusting2003models}.
Though recent efforts have explored AI tools in higher education \cite{luo2025design}, these technologies do not take into account the unique social contexts and outside demands placed on adult learners.
% However, most AI-power educational tools are developed to support learning in K-12 contexts; they are based more in pedagogy than andragogy.

While andragogy characterizes adult learners, pedagogy offers many well-established instructional strategies and cognitive principles that are also applicable to adult learning, such as scaffolding \cite{wood1976role}, formative feedback \cite{shute2008focus}, guided reflection and experimentation \cite{morris2020experiential}, and structured learning activities \cite{bransford2000people}.
These practices are general across both K12 and adult learner contexts.
When applied to adult learning, pedagogical strategies should be adapted to respect learner autonomy, leverage prior experience, and align with goal-oriented motivations \cite{knowles1973adult, tusting2003models}.
In this way, pedagogy is a foundational set of instructional design principles that can be translated to adult learning contexts.

Particularly relevant to adult online learning, \citet{garrison1999critical} introduced the Community of Inquiry (CoI) framework to investigate online learning environments, which suggests that a meaningful educational experience occurs through the interaction of teaching presence, social presence, and cognitive presence.
Cognitive presence is the learner's internal engagement with the learning experience. Social presence is the group collaboration and discussion toward learning outcomes, and allowing learners to project their authentic selves. 
Teaching presence is the organizational structure, design, and facilitation of the learning environment.
% Teaching presence is the organizational structure, design, and integration of the cognitive and social elements that support learning.
% Successful educational experiences require these three elements to work together to engage the learner in a community of inquiry.
% Cognitive, social, and teaching presence are natural components of the K-12 classroom, but special attention should be paid to the construction and organization of a community of inquiry for adult and online education \cite{garrison2010first}.
Social presence is particularly critical in online contexts, where asynchronous and distributed interaction can inhibit the development of interpersonal relationships.
Unlike face-to-face settings where social connections form naturally, online learning environments require deliberate design to foster social presence, as its absence has been linked to learner isolation and disengagement \cite{garrison2007researching}.
The three presences are interconnected: teaching presence creates the conditions for social and cognitive presence to develop and interact, cognitive presence is deepened by collaborative inquiry, and social presence builds motivation and trust for sustained effort.
Adult and online learning environments require intentional design and support for these elements, since adult learners often engage in distributed, asynchronous, and professional learning contexts \cite{meierkord2025trends, garrison2010first}.
The technologies developed by AI-ALOE are designed to support different elements and combinations of elements within the CoI framework.
By looking at the seven tools developed within the institute collectively, we offer a view into how meaningful learning can be supported by an ecosystem of online educational technologies.

Adult learning in online environments involves challenges related to sustained engagement, feedback, and facilitation interaction across distributed and asynchronous contexts.
Artificial intelligence has increasingly been explored as a way to support teaching and learning \cite{9069875, chiu2023systematic}.
Prior research shows that AI-supported educational tools can assist with individualized feedback, content adaptation, and instructional coordination across pedagogical theories \cite{diaz2024artificial}. 
However, empirical studies have also identified several challenges with AI-supported learning, including learner frustration and tools that are difficult to understand or use effectively \cite{10.1145/3597503.3639201, xu2022systematic}. 
In response, several frameworks have been proposed to guide the use of AI in education, including pedagogically oriented classifications \cite{diaz2024artificial} and principle-based approaches focused on transparency, agency, and accountability \cite{fedele2024altai}. 
Despite their contributions, these frameworks often remain high-level and offer limited guidance for translating pedagogical principles into the design of AI-supported tools, particularly in adult learning contexts. 
Efforts to adapt broader educational design frameworks, such as Universal Design for Learning, to AI-supported contexts have faced similar challenges related to complexity and limited empirical support \cite{zhang2024unraveling}.

% REVIEW TYPES OF TOOLS
The tools developed within AI-ALOE span several established categories of educational technology.
Intelligent tutoring systems have been widely studied, with learning gains approaching those of one-on-one human tutoring \cite{vanlehn2011relative}.
Interactive video platforms embed questions and tasks into educational videos, which improves learning outcomes over passive viewing \cite{ploetzner2024effectiveness}.
AI-powered chatbots have grown substantially by applying large language models to provide on-demand instructional support and personalized feedback, though there are ongoing concerns around reliability and academic integrity \cite{labadze2023role}.
Intelligent textbooks also have integrated large language models to generate formative feedback, assessment questions, and interactive dialog grounded in textbook content \cite{sosnovsky2025intelligent}.
Scientific modeling tools allow students to construct, test, and revise hypotheses through interactive simulations \cite{goel2015impact, de2021role}.
Social agents are a more recent innovation, designed to combat student isolation in online learning by suggesting potential collaborators with shared interests and goals \cite{wang2022understanding}.
Since the suite of AI-ALOE tools spans a broad range of techniques, learning goals, and strategies, examining the constraints and impacts of the tools collectively provides a more well-rounded perspective of AI-supported learning tools.

Within human–AI interaction research, design frameworks and guidelines have played an important role in translating high-level principles into actionable support for system design. 
For example, the human-AI interaction guidelines proposed by \cite{amershi2019guidelines} demonstrates how empirical evidence can be consolidated into reusable guidelines that informs evaluation and design practice.
More recent research has emphasized the value of design artifacts like toolkits, cards, and structured resources to support ideation and collaboration when designing AI systems \cite{yildirim2023creating}.
A growing body of work in AI and education has proposed guidelines and frameworks to support the responsible and effective use of AI in learning and assessment \cite{nguyen2025generative, nguyen2025guidelines, nikiforova2025ethical}.
Recent synthesis shows that much of this guidance is framed at the policy level, rather than system guidance \cite{funa2025policy}.
This literature highlights the importance of empirically grounded design guidance that connects pedagogical theory with the practical realities of designing, evaluating, and iterating on AI-powered educational systems for adult learners.

\section{Background}
%\subsection{ National Institute for AI Research on Learning for Online Education and Adult Education}
\subsection{AI-ALOE}
The National AI Institute for Adult Learning for Online Education (AI-ALOE) is a large, interdisciplinary research institute formed through a diverse coalition of research-intensive and regional universities, and community and technical college systems. The institute brings together researchers from artificial intelligence, learning sciences, cognitive science, education, linguistics, and human-computer interaction around a shared mission of improving the quality and accessibility of online and hybrid education for adult learners. AI-ALOE focuses specifically on adults engaged in re-skilling, up-skilling, or continuing education \cite{goel2024ai}---learners whose educational experiences are shaped by limited time, varying prior knowledge, and the need to integrate learning into complex personal and professional lives. In pursuit of this mission, the institute develops and studies AI-driven learning technologies intended to support flexible, self-directed learning pathways, while accounting for the social, cognitive, and institutional constraints that characterize adult education contexts.

We selected AI-ALOE technologies for investigation because they offer a combination of contextual diversity and functional breadth. First, the institute’s technologies span a wide range of adult learning contexts and populations, including rural and urban settings, community colleges and research universities, and disciplines across STEM and the humanities. Second, the portfolio of tools reflects broad coverage of instructional roles and AI capabilities, ranging from tutoring and feedback to social support, metacognitive scaffolding, conceptual modeling, and instructor-facing analytics. This diversity enables analysis across multiple forms of adult learning support. Lastly, these technologies have been examined through sustained, longitudinal studies in authentic adult learning environments, generating rich data from learners, instructors, and design teams over time. Together, these characteristics make AI-ALOE an especially well-suited setting for synthesizing design guidelines that are grounded in practice and sensitive to the realities of adult learning rather than isolated or short-term laboratory studies.

\subsection{AI Technologies}\label{section:ai_technologies}

In this section, we describe each AI technology we examined as part of this study. The technologies were developed by research teams from several academic institutions, including Georgia Institute of Technology, Georgia State University, Harvard University, University of North Carolina at Greensboro, and Vanderbilt University.

\subsubsection{\textbf{Apprentice Tutors}}
A web-based, intelligent tutoring system (ITS) platform designed to support adult learners in practicing domain-specific procedural skills through interactive problem solving \cite{gupta2025beyond}. Tutors can be created in one of two ways: (1) manually, by conducting a cognitive task analysis \cite{lovett1998cognitive} with instructors to inform the design of each tutor, or (2) through a tutor builder component that enables educators to construct interfaces via drag-and-drop interactions and generate expert models with the support of an AI agent \cite{smith2024apprentice}. As learners engage with dynamically generated problems, the system provides real-time correctness feedback, allowing them to iteratively refine their understanding and skills. Instructor- and learner-facing dashboards further support the learning process by visualizing skill mastery and progress over time, helping users monitor performance and make informed instructional or self-regulatory decisions.

\subsubsection{\textbf{iTELL}}
Intelligent Texts for Enhanced Lifelong Learning (iTELL) is an AI-powered platform that helps adult learners practice active and metacognitive reading strategies by paraphrasing and reflecting on key ideas from the material \cite{morris2024automatic}. iTELL ingests instructional text in PDF format and transforms it into an interactive web application that includes content such as short constructed-response questions, page-level summaries, and end-of-page cloze test activities. iTELL's AI module provides formative feedback for learners \cite{morris2023using, morris2025formative} and its analytics dashboard provides actionable insights for educators to aid in instructional decision-making \cite{crossley2025exploratory, crossley2025ai}.

\subsubsection{\textbf{Ivy}}
An interactive video-based AI coaching system embedded in online courses to support procedural skill learning \cite{dass2025ivy,lum2025designing}. Ivy generates structured explanations about ``how'' and ``why'' procedural steps work by integrating symbolic representations of goals and problem decompositions, and states and causal transitions, with LLM-based synthesis of explanations. These representations act as active scaffolds that shape both what is explained and how explanations are organized, supporting adult learners' understanding of complex, multi-step procedures.

\subsubsection{\textbf{Jill Watson}}
A retrieval-augmented generation powered generative-AI-based teaching assistant and cognitive partner that converses with learners using instructor-approved course materials \cite{goel2018jill, kakar2024jill}. It aims to help adult learners access instructional support anytime, anywhere by increasing teacher presence \cite{lindgren2024does} and to deepen cognitive engagement with course content by offering a conversational interface for clarification and reinforcement \cite{maiti2025can}. It supports both retrieval of text and images in response to content-based and logistical queries in both text and audio formats \cite{taneja2024jill}. The system has been deployed across multiple courses in different subjects and has been integrated with existing learning management systems \cite{kakar2024jill}.

\subsubsection{\textbf{SAMI}}
A social agent embedded within online discussion platforms to encourage class participation and social interaction \cite{wang2020jill}. It reads students' posts on the discussion forum, uses LLMs to build knowledge graphs based on student attributes, and uses graph matching to generate recommendations for social connections \cite{kakar2024sami}. It also provides explanations of its reasoning and recommendations. The tool is designed to increase a learner’s sense of social belonging and is used to support emotional well-being and self-efficacy. It aims to help adult learners feel more connected to their peers and instructors.

\subsubsection{\textbf{SMART}}
Student Mental Model Analyzer for Research and Teaching (SMART) provides AI-powered formative feedback on learner-generated summaries and supports iterative revision at a learners’ own pace \cite{deci2000and}. It integrates knowledge-based AI with generative AI to deliver multi-modal feedback, including a visualized knowledge map that highlights key concepts learners have identified or omitted \cite{haddadian2025exploring, kim2024exploring}. Through this AI-augmented summarization process, adult learners are better supported in comprehending core concepts of the material \cite{bae2024study, kim2024investigating}.

\subsubsection{\textbf{VERA}}
A conceptual modeling tool that guides learners through the process of building, simulating, and revising scientific models \cite{an2020scientific,an2025online}. It provides a visual modeling interface connected to a backend simulation engine and is supported by AI-generated coaching and feedback \cite{pmlr-v273-buckley25a}. The tool is used in courses that emphasize systems thinking and inquiry-based learning. It aims to support adult learners by helping them understand and manipulate complex systems through hands-on model construction and experimentation, scaffolding them through any missing domain knowledge they may have \cite{kos2024constructivist,KOS2025SCA}.

%\subsubsection{\textbf{Technology B}}
%A social agent embedded within class communication platforms to encourage participation and interaction. It provides real-time, conversational responses using a combination of language models and knowledge databases. The tool is designed to increase a learner’s sense of social presence and is used to support emotional well-being and self-efficacy. It aims to help adult learners feel more connected to their peers and instructors.

%\subsubsection{\textbf{Technology D}}
%An intelligent tutoring system that gives instructors pre-built, AI-powered tutors for domain-specific skills. It uses symbolic expert systems to support personalized scaffolding and interactive practice. The system includes dashboards for monitoring learner progress and supports subjects like nursing and mathematics. It aims to help adult learners practice structured tasks with feedback and guidance designed to adapt to their skill level.

%\subsubsection{\textbf{Technology E}}
%An interactive video tool that offers real-time prompts and feedback as learners engage with multimedia content. It uses causal and hierarchical representations to diagnose misconceptions and encourage inquiry-based learning. The tool is embedded in courses to support procedural understanding and reinforce problem-solving strategies. It aims to support adult learners by providing structured prompts that encourage reflection and improve understanding of complex processes.

\subsection{AI Technology Deployments}

\begin{figure*}[t]
    \centering
    \includegraphics[width=\textwidth]{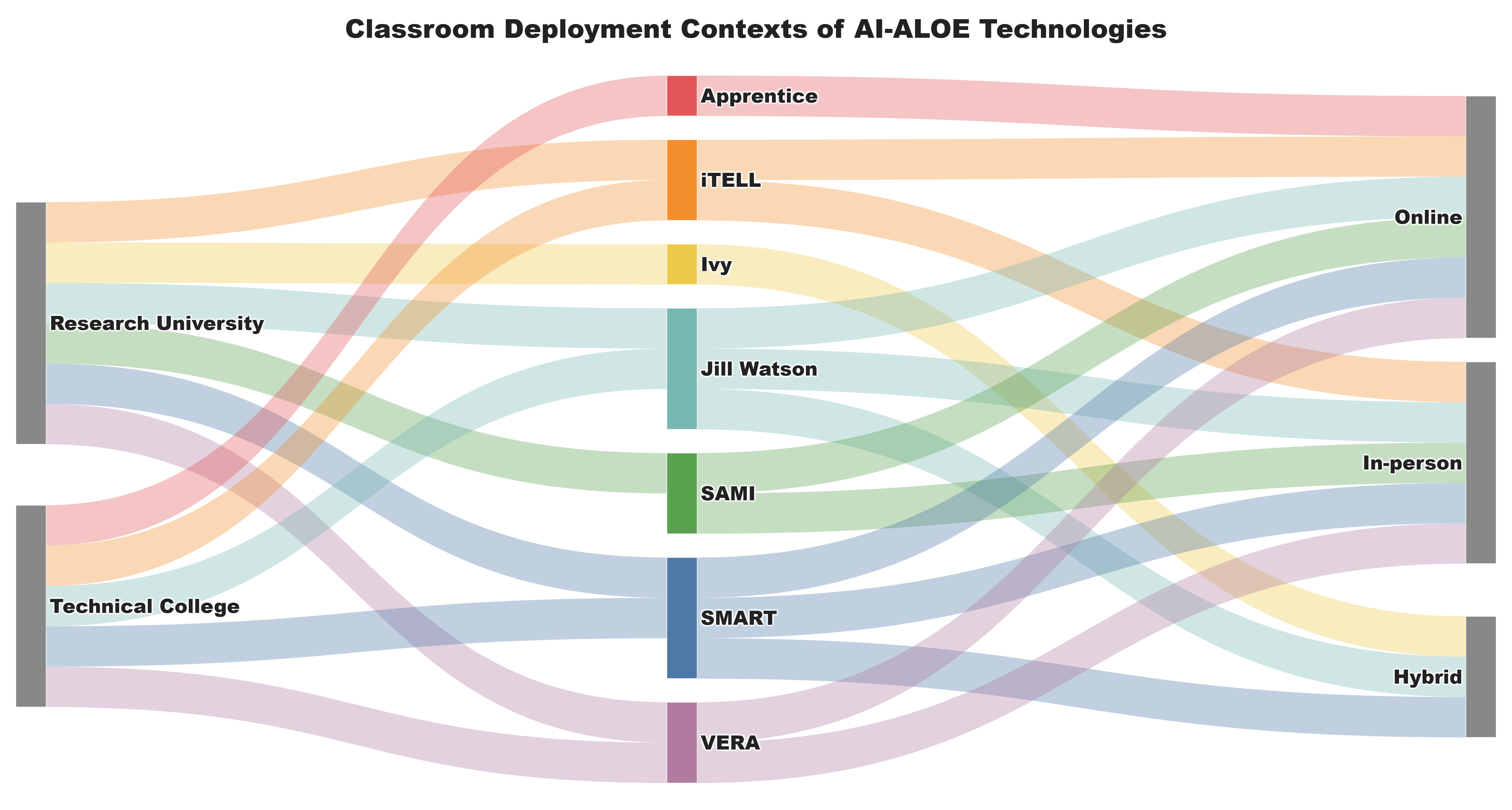}
    \caption{Connections between institution types (research universities and technical colleges), AI-ALOE technologies, and delivery modalities (online, in-person, hybrid).}
    \label{fig:sankey_deployments}
    \Description{A Sankey diagram with three columns. The left column shows two institution types: Research University and Technical College. The middle column shows seven AI-ALOE technologies: Apprentice, iTELL, Ivy, Jill Watson, SAMI, SMART, and VERA. The right column shows three delivery modalities: Online, In-person, and Hybrid. Colored bands flow from institution types to technologies and from technologies to modalities. Research University connects to all seven technologies. Technical College connects to Apprentice, iTELL, Jill Watson, SMART, and VERA. Most technologies connect to both Online and In-person modalities, while Jill Watson and SMART also connect to Hybrid. Ivy connects only to Hybrid. Apprentice connects only to Online.}
\end{figure*}

The AI-ALOE technologies were deployed across a diverse set of adult learning environments, spanning credit-bearing and non-credit courses offered in fully online, hybrid, and in-person formats. These deployments took place across multiple academic institutions and disciplinary contexts, including nursing, algebra, computer science, ecology, web development, and research methods.
Table~\ref{tab:deployment_context} summarizes the subjects, learner age ranges, and geographic contexts of each technology's deployments, and Figure~\ref{fig:sankey_deployments} illustrates how institution types connect to technologies and delivery modalities.

\begin{table*}[t]
\centering
\small
\begin{tabular}{@{} l p{6cm} c p{3.5cm} @{}}
\toprule
\textbf{Technology} & \textbf{Subjects} & \textbf{Age Range} & \textbf{Geographic Context} \\ \midrule
Apprentice   & Algebra, Nursing                                                                              & 21--65    & Online                        \\
iTELL        & Employment Regulations, Intro to Programming, Learning Analytics, Mobile App Dev, NLP, Pharmacology, Research Methods in Psychology & 18--50    & Online, Urban                 \\
Ivy          & Knowledge-based AI                                                                             & 21--65    & Online                        \\
Jill Watson  & Accounting, Computer Science, English, Information Technology, Leadership \& Management, Medicine for Business, Mobile App Dev, Nursing & 20--35    & Online, Rural, Urban          \\
SAMI         & Computer Science                                                                               & 20--50    & Online, Urban                 \\
SMART        & Biology, Computer Science, Education, English, Nursing                                          & $\sim$20s & Online, Urban/Suburban         \\
VERA         & Biology, Cognitive Science, Computer Science, Ecology, Wildlife Management                      & 18--65    & Online, Rural, Urban          \\ \bottomrule
\end{tabular}
\caption{Deployment contexts of AI-ALOE technologies, summarizing the subjects, learner age ranges, and geographic settings in which each technology was used.}
\Description{A table titled “Deployment contexts of AI-ALOE technologies” with four columns: Technology, Subjects, Age Range, and Geographic Context. It lists seven technologies:

Apprentice: subjects include Algebra and Nursing; age range 21 to 65; used online.
iTELL: covers Employment Regulations, Intro to Programming, Learning Analytics, Mobile App Development, Natural Language Processing, Pharmacology, and Research Methods in Psychology; age range 18 to 50; used online in urban contexts.
Ivy: focuses on Knowledge-based AI; age range 21 to 65; used online.
Jill Watson: includes Accounting, Computer Science, English, Information Technology, Leadership and Management, Medicine for Business, Mobile App Development, and Nursing; age range 20 to 35; used online in both rural and urban contexts.
SAMI: focuses on Computer Science; age range 20 to 50; used online in urban contexts.
SMART: includes Biology, Computer Science, Education, English, and Nursing; learners are approximately in their 20s; used online in urban and suburban contexts.
VERA: includes Biology, Cognitive Science, Computer Science, Ecology, and Wildlife Management; age range 18 to 65; used online in rural and urban contexts.

Overall, all technologies are deployed online, with some also spanning urban, suburban, and rural settings, and serving a wide range of subjects and adult learners.}
\label{tab:deployment_context}
\end{table*}

Several deployments focused on supporting social engagement and community building in online courses, where adult learners may experience isolation or reduced opportunities for peer interaction. For example, SAMI was embedded within online discussion platforms in self-paced courses to encourage participation, foster social connection, and support learners’ sense of belonging \cite{kakar2024sami}. In contrast, tools such as VERA and IVY were deployed in classroom-based and hybrid courses to support conceptual understanding and procedural learning through modeling, simulation, and structured explanation \cite{dass2025ivy}. 
Apprentice Tutors was integrated into mathematics and nursing courses to provide targeted, on-demand practice problems with immediate feedback, enabling learners to develop problem mastery outside of limited class time \cite{gupta2024intelligent}. 
iTELL was used in web development and research methods courses to support active reading, comprehension, and metacognitive strategies as learners engaged with complex instructional texts \cite{crossley2025exploratory}. Across deployments, instructional goals included strengthening teaching presence, increasing learner engagement, improving feedback quality, and supporting self-regulated learning.

A defining characteristic of these deployments was the close collaboration between research teams, instructors, and learners throughout design and refinement. Many teams adopted participatory and iterative approaches, working directly with faculty to align system behavior with instructional goals. For instance, the Apprentice Tutors team met regularly with instructors to conduct cognitive task analyses and identify domain-specific tutoring needs, resulting in tutors that were tightly aligned with course content and instructional intent \cite{gupta2024intelligent}. The SMART team introduced automated feedback tools to instructors who were new to AI-assisted instruction, providing structured onboarding and continuing support as instructors gradually integrated AI-generated feedback into their courses \cite{kim2024investigating}. These collaborations allowed instructors to shape how AI features were introduced and framed for learners, while enabling research teams to adapt system designs in response to instructional constraints.

Across multiple deployments, the technologies demonstrated measurable improvements in learner outcomes and stakeholder adoption. Use of Jill Watson was associated with higher average course grades and improved retention compared to control sections \cite{kakar2024jill}. Deployments of SMART showed that learners receiving AI-supported concept feedback produced higher-quality written work \cite{bae2024study}. Apprentice showed higher in-class assessment performance after using the tutor \cite{gupta2025intelligent}. iTELL demonstrated improvements in summary writing quality and sustained engagement across diverse adult learner populations in both hybrid and online contexts \cite{crossley2025ai}. Lastly, results from VERA suggested gains in the complexity of learners’ mental models, though outcomes varied by course structure and instructional emphasis \cite{KOS2025SCA}.

Viewed collectively, the deployments surfaced a set of recurring themes that appeared across technologies, disciplines, and instructional formats. Instructors and learners consistently highlighted similar challenges and points of friction during use, regardless of the specific system involved. The presence of these patterns across otherwise distinct deployments suggested that they reflected broader features of adult learning experiences. These observations motivated the cross-technology analysis presented in this paper, forming the empirical foundation for the design guidelines articulated in the following sections.

\section{Methodology}

To conduct our synthesis, we performed a \textit{reflexive thematic analysis} \cite{braun2006using, braun2019reflecting} of user and stakeholder feedback. The goal was to support future development of AI-powered education tools by highlighting common design considerations, user needs, and contextual factors that emerged across our multiple technologies and deployments. This section details how we compiled the relevant data and performed the analysis. 

\subsection{Data Source Collection}
We began our analysis by assembling a corpus of qualitative data that captured how the AI-ALOE technologies were designed, deployed, and experienced in  adult learning contexts. These data were drawn from multiple sources to reflect both the intentions of research teams and the usage experiences of learners and instructors. Specifically, we collected a total of 15 presentations from the research and development teams of the technologies that documented design rationales, relevant learning science frameworks, and instructional considerations; transcripts from 17 focus groups conducted with learners, instructors, and administrators; and 3 sets of cross-team feedback from NSF program committee meetings.

To support a systematic, cross-technology analysis, we transformed these heterogeneous materials into a standardized data format. From presentations, reports, and design documents, we extracted discrete statements reflecting educational concepts such as learning science theories and design considerations articulated by research teams. From focus group transcripts, we extracted individual thoughts, experiences, and reflections expressed by learners, instructors, and administrators, preserving the original phrasing to maintain context. Each extracted item was entered as a single row in a digital spreadsheet and annotated with metadata including the associated technology, the raw text excerpt, and the source context (e.g., ``SMART - Instructor Focus Group'', ``iTELL -  Design Presentation'', etc.). This standardized representation enabled us to assign unique identifiers to each data item (e.g., ``SMART-Learner Focus Group-299''), allowing insights to be traced back to their original source throughout analysis if additional context was required. This structured dataset formed the foundation for the reflexive thematic analysis described in the following section. 

\subsection{Reflexive Thematic Analysis}\label{section:thematic_analysis}

\begin{figure*}
    \centering
    \includegraphics[width=1\linewidth]{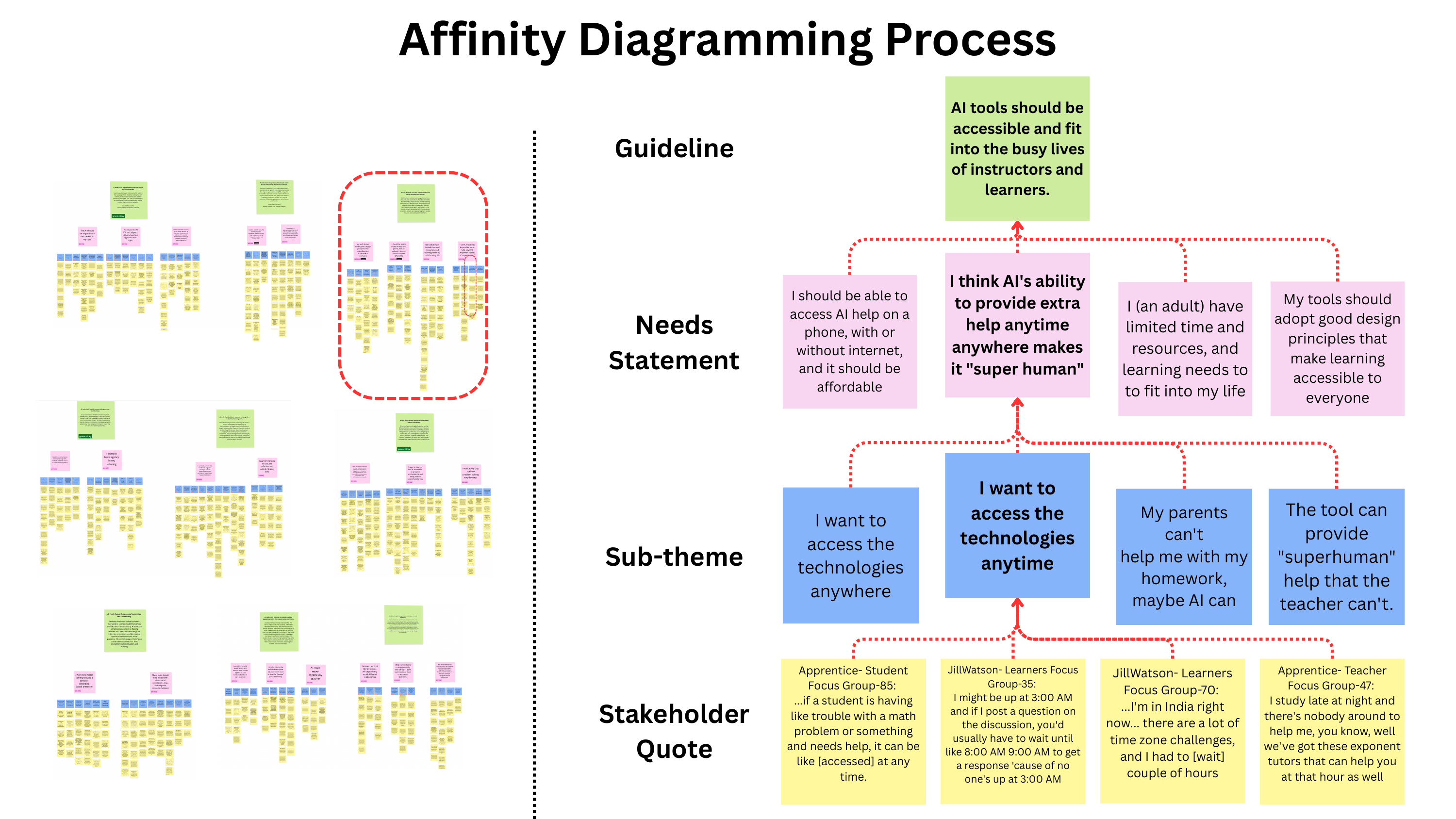}
    \caption{Affinity diagramming process used to derive design guidelines from qualitative data. 
    Stakeholder quotes (yellow notes) were clustered into sub-themes (blue notes) capturing recurring perspectives. 
    Sub-themes were synthesized into needs statements (pink notes) spanning multiple stakeholder groups. 
    Finally, needs statements informed distilled design guidelines (green notes).}
    \label{fig:affinity_diagram}
    \Description{
    The figure presents an affinity diagram used to organize qualitative data from focus groups and interviews. 
    At the base level, individual data items (yellow sticky notes) contain direct quotes, observations, or insights from participants (e.g., learners describing feelings of isolation or teachers expressing frustration with interface design). 
    These yellow notes are clustered under blue sticky notes, which summarize shared patterns in the first person perspective (e.g., “I want to feel a sense of community”).

    Next, the blue notes are grouped under higher-level pink sticky notes, which articulate broader thematic concerns across stakeholders (e.g., “I want to design AI tools that are intuitive, easy to use, and promote natural interactions”). 
    Finally, at the highest level, green sticky notes represent distilled design guidelines framed as “AI tools should…” statements (e.g., “AI tools should foster social connection and community”). 
    The diagram visually traces the process from raw participant data to thematic categories and, ultimately, to design guidelines that capture the needs and values of diverse stakeholders.}
\end{figure*}

We conducted a reflexive thematic analysis \cite{braun2006using, braun2019reflecting} to synthesize patterns across the collected data. The analysis was carried out by a deliberately diverse research team that included at least one representative from each of the technology design teams, one researcher who led and conducted the focus groups, and 3 additional HCI researchers who were not part of the development or focus group teams. To ensure a comprehensive overview, all team members participated in a collaborative synthesis process to analyze all of the data following the procedure of Holtzblatt \cite{holtzblatt1997contextual}. Team members represented over five institutions and brought complementary perspectives from computer and learning sciences, educational technology, and human-computer interaction. This composition was intended to ensure a plurality of perspectives informed the analysis, spanning design intent, deployment experience, and learner- and instructor-facing observations. Through iterative discussion and debate, the team collaboratively identified, refined, and synthesized themes, drawing on both the data and their situated expertise with adult learning technologies. As a result, we acknowledge that it is possible another research team, drawing on different experiences and values, might arrive at an alternate set of themes or guidelines when analyzing the same dataset. Indeed,  we selected this approach because the goal of this study was not to establish a single objective account of stakeholder experiences, but to surface recurring design considerations grounded in the situated experiences of learners, instructors, and technology developers across multiple AI systems and contexts.

All extracted data items---approximately 1,600 individual statements---were imported into Miro\footnote{https://miro.com/} and initially organized by technology. Each item was represented as a yellow sticky note containing a verbatim quote, observation, or design insight from the dataset. Following an affinity diagramming process, the team collaboratively grouped related yellow notes into first-level thematic clusters, each labeled with a blue sticky note (See Figure~\ref{fig:affinity_diagram}). These blue notes captured shared patterns across small sets of related items (typically four to six) and were framed in the first-person (``I'') perspective to reflect the dominant stakeholder voice represented in the underlying data (e.g., learner, instructor, or developer). For example, a blue note such as ``\textit{I want students to develop transferable critical thinking skills}'' reflected a recurring instructor-oriented concern present across multiple data sources.

Next, the team grouped related blue notes into higher-level pink notes, also written in the ``I'' perspective, but reframing broader concerns of multiple stakeholders as desired features. For instance, a pink theme such as ``\textit{I want to design AI tools that are intuitive, easy to use, and promote natural interactions to avoid confusion over the interface and expectations of the tool}'' encompassed learner anxieties about data deletion ("\textit{I'm afraid of accidentally deleting my work}"), instructor concerns about usability ("\textit{I make mistakes because of the design of the UI}"), and developer reflections on overall interface design. Throughout this process, themes were iteratively discussed, merged, split, or reframed as the team revisited the data and reflected on alternative interpretations.

Finally, the pink themes were synthesized into a set of design guidelines, represented as green sticky notes and framed as actionable statements beginning with ``AI tools should...''. This synthesis involved a collaborative review where the entire research team examined each thematic group of pink notes, proposed candidate guidelines, and iteratively revised the wording until reaching a consensus that each guideline accurately reflected the consolidated stakeholder voice. Each guideline was accompanied by a detailed description and further annotated along several dimensions, including the primary stakeholders affected, the types of technology problems addressed (e.g., access, trust, engagement), and the Community of Inquiry dimensions---cognitive, teaching, and social presence \cite{garrison1999critical}---that the guideline supports. These annotations supported later analysis and application of the guidelines, including the heuristic evaluation described in Section 6.

% TEMPORARILY COMMENTING OUT DIAGRAM BECAUSE ITS TOO BIG FOR THE PREVIEW
% \begin{figure}
%     \centering
%     \includegraphics[width=1\linewidth]{images/temp_affinity.png}
%     \caption{TEMP AFFINITY DIAGRAM, CHANGE BEFORE SUBMITTING}
%     \label{fig:affinity_diagram}
% \end{figure}

\begin{figure*}[h]
    \centering
    \includegraphics[width=1\linewidth]{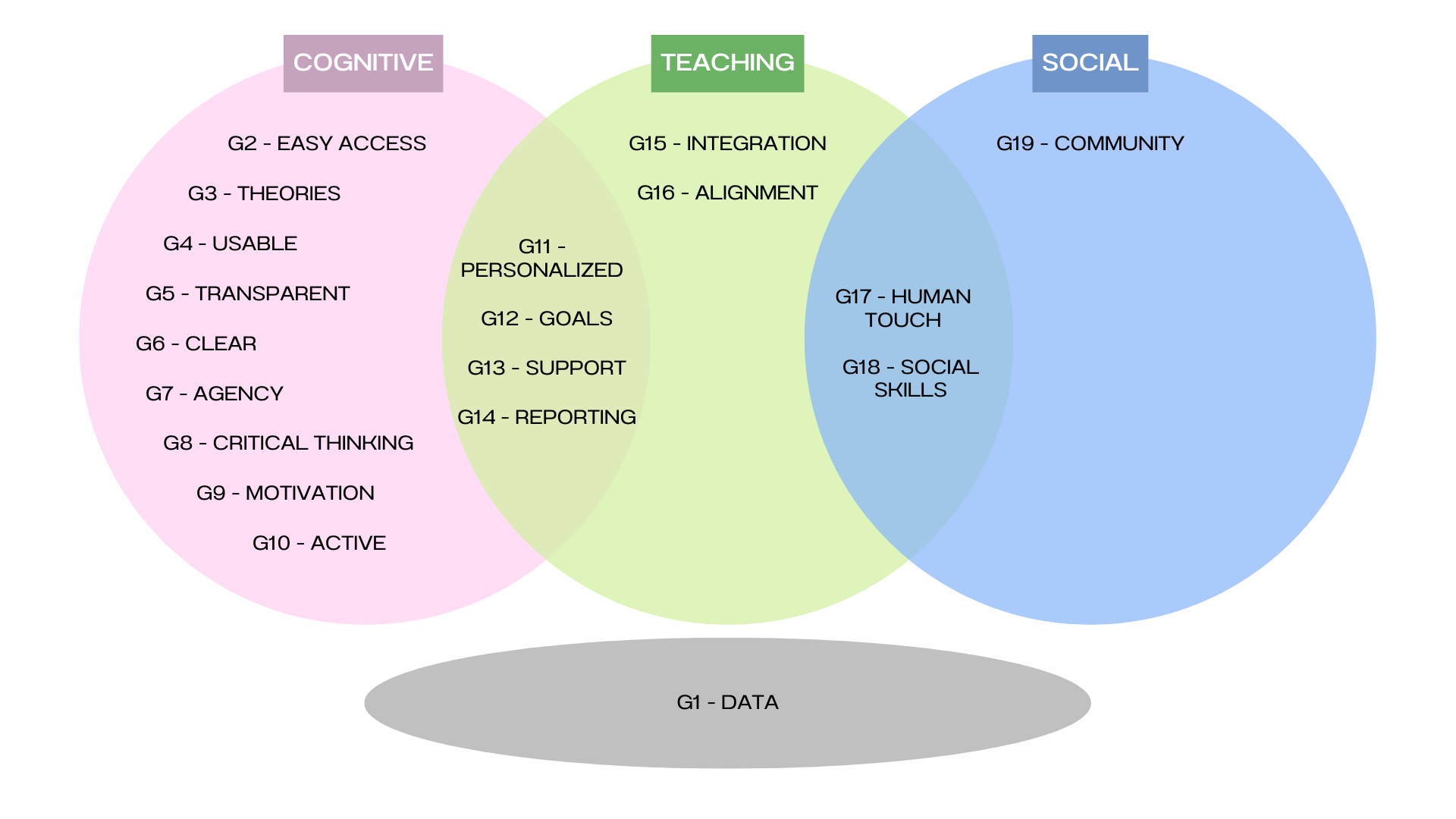}
    \caption{Design guidelines organized across three overlapping Community of Inquiry dimensions: Cognitive (learner experience and understanding), Teaching (instructional design and alignment), and Social (community and interpersonal development). Shared regions indicate principles that bridge domains, while G1 (Data Practices) sits outside the CoI framework.}
    \label{fig:CoI_alignment_venn}
    \Description{
    The figure shows a conceptual framework of design guidelines organized across three overlapping Community of Inquiry (CoI) dimensions: Cognitive, Teaching, and Social. 
    The Cognitive domain represents learner experience and understanding, the Teaching domain represents instructional design and alignment, and the Social domain represents community and interpersonal development.
    Individual guidelines (labeled G2-G19) are placed within the domains they primarily address.
    Guidelines located in overlapping regions indicate principles that bridge multiple CoI dimensions, relating to two of the cognitive, instructional, or social aspects of learning.
    Below the three domains, an oval labeled G1 (Data Practices) sits outside the CoI framework, since is not specific to any single CoI category. 
    }
\end{figure*}

\begin{table*}[]
\begin{tabular}{|l|p{0.4\linewidth}|p{0.1\linewidth}|p{0.1\linewidth}|p{0.2\linewidth}|}
\hline
ID  & Guideline                                                                                               & CoI                 & Stakeholder                    & Problem                             \\ \hline
G1  & AI tools should be open about data practices and share governance with learners.                        & -                   & Developer, Researcher, Learner & Access, Trust                       \\ \hline
G2  & AI tools should be accessible and fit into the busy lives of instructors and learners.                  & Cognitive           & Teacher, Learner               & Access                              \\ \hline
G3  & AI tools should be informed by learning science and learning theories.                                  & Cognitive           & Researcher, Developer          & Low Learning                        \\ \hline
G4  & AI tools should be easy to understand and frictionless to use.                                          & Cognitive           & Teacher, Learner               & Low Engagement, Access              \\ \hline
G5  & AI tools should be transparent and explainable.                                                         & Cognitive           & Learner                        & Trust                               \\ \hline
G6  & AI tools should minimize cognitive load by presenting essential information clearly and simply.         & Cognitive           & Developer                      & Low Learning, Low Engagement        \\ \hline
G7  & AI tools should provide learners with agency over their learning.                                       & Cognitive           & Learner                        & Low Adoption, Low Engagement        \\ \hline
G8  & AI tools should cultivate learners' metacognitive and critical thinking skills.                         & Cognitive           & Learner, Researcher            & Low Learning                        \\ \hline
G9  & AI tools should support learners’ motivation and cultivate self-efficacy.                               & Cognitive           & Learner                        & Low Engagement                      \\ \hline
G10 & AI tools should encourage active, constructive, and interactive engagement over passive consumption.    & Cognitive           & Researcher, Developer          & Low Learning, Low Engagement        \\ \hline
G11 & AI tools should personalize learning based on learners' knowledge, skills, and abilities.               & Cognitive, Teaching & Teacher, Developer             & Low Learning                        \\ \hline
G12 & AI tools should offer meaningful challenges that are aligned with learners' context, goals, and career. & Cognitive, Teaching & Learner                        & Low Engagement                      \\ \hline
G13 & AI tools should provide accurate, contextualized help when learners need it.                            & Teaching, Cognitive & Developer                      & Low Engagement, Trust, Low Learning \\ \hline
G14 & AI tools should provide actionable insights to instructors, learners, and researchers.                  & Teaching, Cognitive & Teacher, Researcher, Learner   & Low Engagement, Low Learning        \\ \hline
G15 & AI tools should integrate with users' existing educational technology ecosystem.                        & Teaching            & Developer                      & Low Adoption                        \\ \hline
G16 & AI tools should align with instructional practices and course content.                                  & Teaching            & Teacher                        & Low Adoption, Trust                 \\ \hline
G17 & AI tools should maintain the human touch and supplement rather than replace human instructors.          & Social, Teaching    & Learner, Teacher, Developer    & Low Adoption, Trust                 \\ \hline
G18 & AI tools should scaffold and support learners in developing their social competencies.                  & Social, Teaching    & Learner, Developer             & Low Engagement                      \\ \hline
G19 & AI tools should foster social connection and community.                                                 & Social              & Learner, Developer             & Low Adoption, Low Engagement        \\ \hline
\end{tabular}
\caption{The 19 design guidelines for AI in adult learning technologies. 
Each guideline is mapped to the Community of Inquiry dimension it supports, the primary stakeholders it addresses, and the technology problems it seeks to mitigate.}
\Description{A table titled “The 19 design guidelines for AI in adult learning technologies” with five columns: ID, Guideline, Community of Inquiry (CoI) dimension, Stakeholder, and Problem addressed.

It lists 19 guidelines:

G1: Transparency about data practices and shared governance; no CoI category; stakeholders are developers, researchers, and learners; addresses access and trust.
G2: Accessibility and fit for busy users; cognitive; teachers and learners; addresses access.
G3: Grounding in learning science and theory; cognitive; researchers and developers; addresses low learning.
G4: Ease of use and low friction; cognitive; teachers and learners; addresses low engagement and access.
G5: Transparency and explainability; cognitive; learners; addresses trust.
G6: Minimizing cognitive load with clear information; cognitive; developers; addresses low learning and engagement.
G7: Supporting learner agency; cognitive; learners; addresses low adoption and engagement.
G8: Developing metacognitive and critical thinking skills; cognitive; learners and researchers; addresses low learning.
G9: Supporting motivation and self-efficacy; cognitive; learners; addresses low engagement.
G10: Promoting active and interactive engagement; cognitive; researchers and developers; addresses low learning and engagement.
G11: Personalizing learning; cognitive and teaching; teachers and developers; addresses low learning.
G12: Providing meaningful, context-aligned challenges; cognitive and teaching; learners; addresses low engagement.
G13: Offering accurate, contextualized help; teaching and cognitive; developers; addresses low engagement, trust, and learning.
G14: Providing actionable insights; teaching and cognitive; teachers, researchers, and learners; addresses low engagement and learning.
G15: Integrating with existing educational technology; teaching; developers; addresses low adoption.
G16: Aligning with instructional practices and content; teaching; teachers; addresses low adoption and trust.
G17: Maintaining human involvement alongside AI; social and teaching; learners, teachers, and developers; addresses low adoption and trust.
G18: Supporting development of social competencies; social and teaching; learners and developers; addresses low engagement.
G19: Fostering social connection and community; social; learners and developers; addresses low adoption and engagement.

Overall, most guidelines emphasize the cognitive dimension, with additional focus on teaching and social aspects, targeting learners, teachers, developers, and researchers, and addressing issues such as low learning, engagement, adoption, access, and trust.}
\label{tab:guidelines}
\end{table*}

\section{Design Guidelines}

% OTHER THINGS WE ALSO INCLUDE IN THE TABLE?????

% \subsection{Design Guidelines} 

Table~\ref{tab:guidelines} shows the 19 guidelines that resulted from our analysis. Each guideline represents a systematic distillation of the multi-level thematic groupings described in our methodology, moving from specific stakeholder concerns to higher-order design principles. By grounding each guideline in the recurring needs and friction points surfaced across various adult learning contexts, we ensure they remain actionable and relevant to the practical challenges of educational technology design.
As described earlier, AI-ALOE adopted \citet{garrison1999critical}'s Community of Inquiry as a core framework to guide the design of its learning tools.
Accordingly, these guidelines are labeled according to the associated Community of Inquiry element.
% We will discuss a few representative examples from each category and present further descriptions of each guideline in the supplementary materials. 
Each guideline is also labeled by the stakeholder most directly affected if it is not satisfied, as well as by the potential technological issues that could be mitigated through its application.
Figure~\ref{fig:CoI_alignment_venn} illustrates how the guidelines overlap with Community of Inquiry dimensions. 
In the supplementary materials, we provide the full affinity diagram, including the stakeholder needs statements and supporting themes behind each guideline.

\begin{figure*}[h]
    \centering
    \includegraphics[width=1\linewidth]{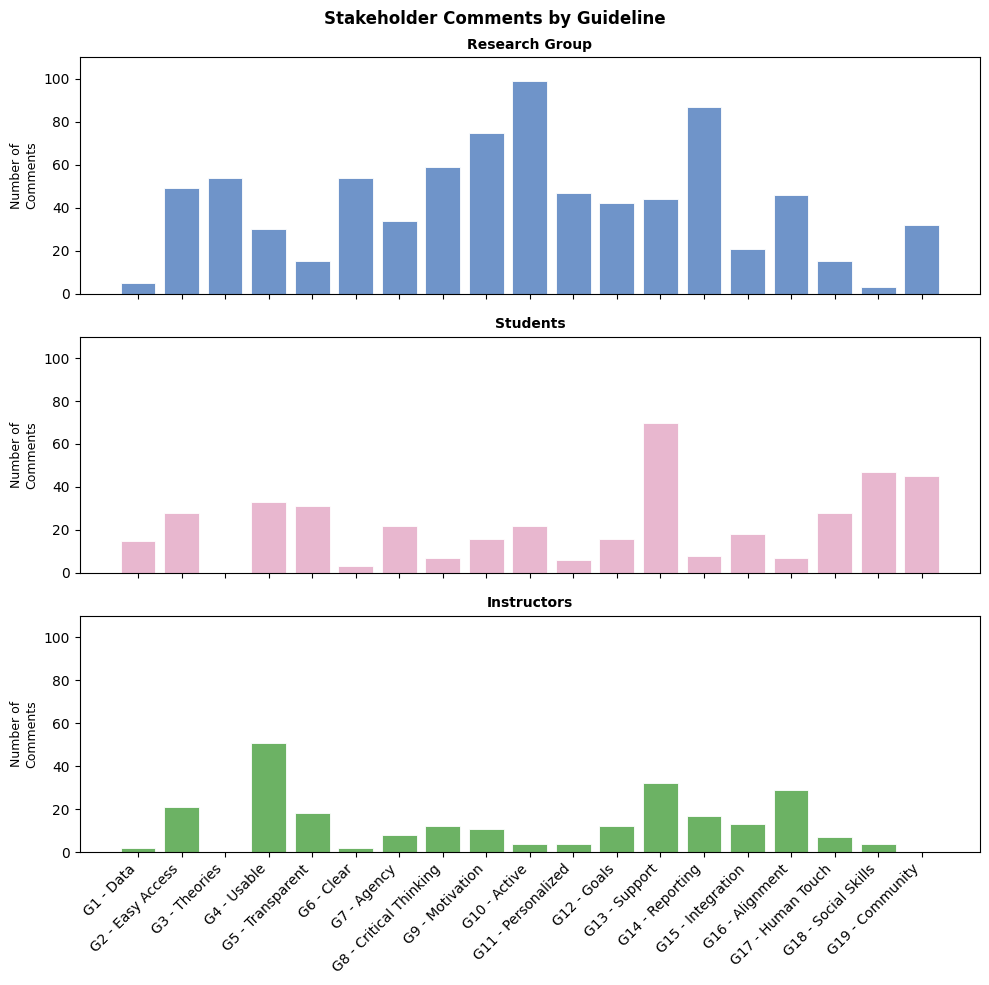}
    \caption{Number of stakeholder comments by design guideline, broken down by group (research team, students, and instructors). Each subplot shows the total volume of comments for each guideline according to the subgroup.}
    \label{fig:stakeholder_comments_distribution}
    \Description{
    The figure show three subplot bar charts with the number of comments provided by three stakeholder groups (research group, students, and instructors) across 19 design guidelines (G1-G19).
    The top subplot (Research Group, blue) shows that the research team contributed the most comments overall. Guidelines G10 (Active), and G14 (Reporting) recieved the highest counts, while G1 (Data) and G18 (Social Skills) received the fewest. Notably, researchers were the only stakeholder group to mention G3 (Theories).
    The middle subplot (Students, pink) shows that students were most concerned with G13 (Support), G18 (Social Skills), and G19 (Community). Relative to the other groups, students were more concerned about social interaction and community, and less concerned about personalized learning, and clear visual presentation.
    The bottom subplot (Instructors, green) shows that teachers commented most on G4 (Usable), G13 (Support), and G16 (Alignment). Teachers did not comment on G19 (Community), and had few comments on G1 (Data), G6 (Clear), G10 (Active), G11 (Personalized), and G18 (Social Skills).
    }
\end{figure*}

We only labeled the stakeholders who would experience an immediate impact if the guideline is not addressed.
For example, guideline G4 says that ``AI tools should be easy to understand and frictionless to use''.
Making a tool easy to understand will be an immediate benefit to the end users, the learners and instructors.
If the tool is not easy to understand, users may stop engaging with the tool.
Though the researcher is not immediately impacted by this issue, they will be affected in the future when they are unable to collect research data due to low engagement and adoption.
We chose not to include the researcher in the stakeholders because, with downstream effects, nearly every stakeholder could be impacted by every guideline which would lead to a trivial taxonomy.
To preserve the usefulness and clarity of the guidelines, we labeled each guideline with only the immediately impacted stakeholders.

In our supplementary materials, we also present the low-level themes and mid-level themes that support each design guideline in the full affinity diagram.
We believe that by reviewing the themes surfaced from concerns of stakeholders, future researchers and developers can generate practical solutions to meet the guidelines based on specific real-world concerns.
In addition, reviewing the supporting themes could help apply the guidelines to broader contexts, beyond the ones from which they were derived.

Across the 19 guidelines, we observed clear differences in how concerns were distributed among stakeholder groups (Figure~\ref{fig:stakeholder_comments_distribution}).
Instructors most frequently commented on the usability of AI tools (G4) and students most frequently commented on the amount of educational support provided (G13), but both stakeholders commented heavily on these two dimensions. 
Instructors also frequently highlighted the alignment of AI tools with their personal instructional approach (G16) because using AI tools with contrasting approaches can be a challenge.
Student comments were also particularly concentrated on guidelines addressing building community (G19) and social competencies (G18), indicating a strong concern for how AI will shape peer and professor interaction.
In contrast, researcher comments were more prominent in guidelines tied to instructional best practices (G8, G9, G10) and using data for learning engineering (G14).
The category of Learning Theories (G3) was also only a concern for researchers and not mentioned by the other stakeholders.
While several guidelines received equal attention from all three groups (G2, G4, G5, G13), the uneven distribution demonstrates that different stakeholders prioritize different dimensions of AI-supported learning.
The guidelines collectively provide a broad representation of the concerns raised across all stakeholder groups.

Guideline \textbf{G1 (Data Practices)} was not assigned a Community of Inquiry component. 
This guideline was derived mainly from student concerns about not knowing what data is being collected, how it is stored, and who can access it.
Since aligning AI technologies with G1 would not increase student cognitive, social, or teaching presence, we elected not to force G1 into the Community of Inquiry framework.
Students expressed a desire to have meaningful control over their data, to review, manage, or delete it. 
% (\textit{``We don't really know how to interact with the data that we gave [the tool] and how to change or delete [the data].''})
AI technologies that use third-party Large-Language Models need to be transparent with users in an educational setting on what data is being sent, who has access to their data, and what they are using their data for
(\textit{``If I have to share some sensitive data, where is the data going to? Who has access to it? So that's completely unknown currently. That information [is] not available anywhere. And moving forward as we are integrating chatGPT with it, it'll be going to again a third party vendor.''}).
Higher education students have an implicit trust in their institution to handle data ethically \cite{slade2019learning}, but that trust does not extend to data collected by a third-party educational tool \cite{park2021college}.
Clear communication about data use can build trust, reduce user anxiety, and empower students as active stakeholders in how their information is being handled \cite{mutimukwe2022students}.

\subsection{Cognitive Presence}

Sustained engagement in an educational experience allows learners to construct meaning, and is a core component of cognitive presence. 
Adult students and instructors need technologies that allow them to engage in educational experiences despite the constraints of their daily responsibilities.
\textbf{Accessibility and flexibility (G2)}  emerged as concerns for adults, as both learners and instructors juggle limited time, resources, and devices. 
% Students want tools that support their learning disabilities (\textit{``I have a learning disability called Dyscalculia. And the unfortunate bit is the AI tutor doesn't seem to take that into consideration.''}) and that they can access on more convenient devices (\textit{``I feel like it would've been much more beneficial, um, for me to have been able to access it on my mobile device.''}).
Students emphasized the need for tools that accommodate learning differences, are mobile-friendly, and are usable without strong internet access. These features let adults fit learning into their busy lives instead of fitting their life around learning.
% (\textit{``I think it's designed with the working professional in mind. So, a lot of people just don't have extra time.''}).
% Learners also discussed how they have limited time and resources, so learning needs to fit into their life, instead of fitting their life around learning (\textit{``I think it's designed with the working professional in mind. So, a lot of people just don't have extra time.''}).
% Teachers also cite that their class time is limited (\textit{``I don't delve deeply into simplifying radicals because I don't have the time. If [the students] had somewhere where they could go practice on simplifying radicals, that would be great.''}), so AI tools can provide anytime anywhere extra help (\textit{``I might be up at 3:00 AM and if I post a question on the discussion board but you'd usually have to wait until like 8:00 AM 9:00 AM to get a response `cause of no one's up at 3:00 AM''}).
Instructors similarly reported limited class time and highlighted the value of AI for providing on-demand support, especially outside normal hours (\textit{``I might be up at 3:00 AM and if I post a question on the discussion board but you'd usually have to wait until like 8:00 AM 9:00 AM to get a response `cause no one's up at 3:00 AM''}).
By aligning with universal design principles to make technology accessible, AI tools can make learning more flexible, inclusive, and sustainable for everyone.

Building technologies with a strong \textbf{Theoretical Foundation (G3)} was a concern voiced entirely by the research teams.
The development teams behind the AI technologies aimed to ground their tools in established theories,  such as self-determination theory---to maintain motivation through autonomy, competence, and relatedness \cite{deci2013intrinsic}---or multimedia learning theory---to improve engagement and retention by using multiple representations of content \cite{mayer1997multimedia}.
Although other stakeholder groups mentioned motivation, critical thinking, or active engagement, instructors and students never explicitly discussed the importance of  theoretical or empirical evidence for the effectiveness of an educational intervention.
Nevertheless, AI technologies should be grounded in evidence to ensure they effectively support desirable outcomes, such as comprehension, motivation, and knowledge construction.

Learning already demands substantial cognitive effort, making it important for learning technologies themselves to be easy to use.
Guideline \textbf{G4 (Frictionless)} strives to make AI tools intuitive, so students and instructors can focus on learning.
Brief, just-in-time onboarding (e.g., tooltips, short videos, examples) should supplement the user interface to clarify expectations and common actions
(\textit{``As funny as that might sound, whether there be a video, maybe like an instructional video, just more information on it because I think there was some confusion on how to actually use [the tutor].''}). 
Adults need brief, contextual, and action-oriented onboarding to support confidence and rapid sense-making \cite{strahm2018generating}.
Importantly, this guideline addresses usability-related friction rather than eliminating productive struggle that supports learning.
Reducing friction in how a tool is accessed and operated builds confidence and trust in the AI technology.
%(\textit{``When we feel insecure about the tutors, we do not want to share this with students''}).
According to the Technology Acceptance Model, perceived ease of use is a primary predictor of users’ confidence, trust, and willingness to adopt new technologies \cite{marangunic2015technology}.

Students heavily commented on the need for \textbf{Transparency (G5)} in AI learning tools.
Learners wanted explanations for why they received the guidance they did, especially when AI tools tell them they are wrong.
% (\textit{``It really helped me, there was even the little scheme that made me understand what I did right or wrong''}).
The tools should provide support for learners to construct mental models for how the tool operates
(\textit{``I would like to see the inner workings or...what are the data points that essentially are being used to make these recommendations.''}).
Prior work on explainable AI in education shows that explanations enable learners to understand how AI systems work, judge the credibility of their outputs, and regulate their own learning \cite{KHOSRAVI2022100074}.
Encouraging self-regulation is essential for adult and self-directed learners \cite{merriam2001andragogy}.

Along with designing tools that are backed by learning science theories, researchers also stressed the importance of designing for \textbf{Cognitive Load Management (G6)}.
Learners benefit when tools reduce distractions and focus attention on what matters most.
Irrelevant or incongruent information places unnecessary demands on working memory and can impair comprehension, problem solving, and memory, particularly for older adults who are more susceptible to distraction \cite{weeks2014disruptive}.
Tools that align design features to instructional content and assessment can reduce extraneous demands on working memory to allow more cognitive resources directed toward learning.
At the same time, certain design elements, like interactivity in digital environments, may introduce extraneous demands while still enhancing engagement and learning when they are aligned with the learning goals \cite{skulmowski2022understanding}.
By carefully crafting a digital learning environment, AI tools can reduce extraneous demands, improve engagement and motivation, and ultimately increase adoption.

A key characteristic of adults, in contrast to children, is they have greater \textbf{Agency (G7)} in their learning \cite{merriam2001andragogy}. 
AI tools should offer choices in how they engage with content and control over how AI supports them. 
Students consistently voiced a desire for control over how the tool provides support, resources, and content
(\textit{``I think just in general, the more options you have for any type of support is always good.''}).
Learners should also be provided with guidance to scaffold their ability to make effective choices.
This could be done through actionable insights and visualizations (\textbf{G14}), contextualized help (\textbf{G13}), or developing metacognitive strategies (\textbf{G8}).
By fostering autonomy and self-direction, AI tools not only reduce barriers to adoption but also strengthen motivation, ownership, and long-term learning outcomes.

Beyond delivering answers, AI should guide learners in using \textbf{Metacognitive Strategies (G8)} such as summarization, self-explanation, and reflection to deepen understanding. 
Tools can also coach students to make thoughtful choices about how they learn—helping them monitor progress, compare approaches, and practice higher-order reasoning. 
Stakeholders stressed the importance of not just including reflective prompts, but scaffold how to engage in reflection
(\textit{``AI is telling you how to reflect, [but is it] giving you the tools to do so?''}).
By fostering reflection and critical thinking, AI supports not only immediate task success but also transferable skills for lifelong learning.

\textbf{Motivation and Self-efficacy (G9)} were also concepts that stakeholders named as important for cultivating better learning.
Researchers wanted to motivate students through in-app messaging, 
% (\textit{``Offer motivational messages and constructive suggestions''})
competition, 
% (\textit{``Leaderboard helps to build social presence by allowing users to see how they stack up against their peers.''} and \textit{``Healthy peer competition can encourage students to stay engaged and supports self-regulated learning''})
and rewards.
% (\textit{``Healthy peer competition can encourage students to stay engaged and supports self-regulated learning''}).
% (\textit{``Motives for learning mix of intrinsic and extrinsic rewards''}).
Students and teachers wanted to make progress more apparent in the tools 
% (\textit{``I think it might help students too, you know, I think students might benefit from seeing progress.''}) 
but in a way that encourages students to keep trying 
(\textit{``You know, you watch that little percentage bar not increase or worse, decrease. I could see that being problematic for some people, like causing frustrations or discouraging them.''}).
Teachers also voiced the need for step-by-step support 
% (\textit{``I feel I need steps for things. ... How do I get to that point of knowing what A equals?''}) 
so students continue to make progress.
% (\textit{``Sometimes progress is you get a D, you take a class over again and then you get a C. Sometimes the progress is slow.''}).
Effective AI tools should cultivate learners' motivation and belief in their ability to succeed.

One of the most commented on topics was the desire \textbf{Actively Engage (G10)} students in their learning.
Tools designed around the ICAP framework can scaffold inquiry, generative tasks, and peer interaction to strengthen cognitive presence \cite{chi2014icap}. 
Learners feel and stay more engaged when prompted to question, explain, and collaborate rather than just consume information
(\textit{``I do think that the value is feeling connected and having it feel like you're actively participating in like class rather than feeling like I'm just watching a bunch of videos and there's no office hours or there's no one to talk to.''}).
This active engagement helps learners build deeper understanding and retain knowledge more effectively.

\subsection{Teaching Presence}

Within CoI, the organizational structure of the learning experience is described as teaching presence. 
AI tools can support teachers in their organization by \textbf{Integrating (G15)} easily into systems already in use, such as learning management systems.
Integration should follow open standards to avoid vendor lock-in, support interoperability, and reduce cost. 
Effective integration makes the tool feel like a natural extension of the existing ecosystem rather than an added burden.
% (\textit{``So I think a more single integrated point of contact will be much better for my personal experience. Uh, so yeah, I mean this integrated experience instead of distributed experience right now that will help me better interact and learn in this whole course''}).

\textbf{Course Alignment (G16)}  was a consistent concern from instructors. 
Teachers repeatedly voiced that they wanted the tools to support their specific content and practices
% (\textit{``Having that domain-specific knowledge is pretty important.''}).
(\textit{``I don't know about other styles, in my style there will only be 1 choice. Not many choices like in this box.''}). 
% Teachers also stressed the need to feel comfortable using the tools themselves before introducing them to students.
% (\textit{``We need to be able to use it before we can start to grade it for the students.''})
Even when teaching the same material, different instructors may structure lessons, sequence activities, or emphasize concepts in distinct ways.
Tools with fixed instructional workflows or rigid content structures are unlikely to transfer between different instructors.
Without clear alignment to instructional practices and course content, teachers are unlikely to adopt AI tools in their classrooms, regardless of potential benefits or technical quality \cite{mills2019alignment}.
For educational AI tools to be successful, they need to earn the trust and support of instructors before the tool is used with students.

\subsection{Cognitive and Teaching Presence}

Many guidelines have the potential to support multiple forms of presence at once. 
Within the Community of Inquiry framework, the presences are deeply interconnected with one another and this interconnection is necessary to sustain meaningful learning.
Some guidelines naturally extend their influence across multiple dimensions of presence rather than being confined to a single category.

\textbf{Personalized Learning (G11)} was described as essential for adult learners, who may have incredibly varied prior education and experience.
Teachers noted that personalization can help bridge gaps for those balancing work, family, and education, by ensuring that time spent on learning is both efficient and relevant 
(\textit{``It would really help bridge the gap for a lot of people, particularly [...] people with kids, people with full time jobs... who might not otherwise be able to get instruction in person.''}).
% Students also wanted adaptive learning sequences to make the most out of their time 
% (\textit{``I might want to drill down more on certain topics, and at the same time something that helps me calibrate, okay, this is what I need to excel the class... if some tool can help me plan, then that will suffice my learning needs.''}).
By engaging the students through tailored learning experiences, this guideline seeks to increase cognitive and teaching presence.
Personalized learning pathways are essential for keeping adult learners engaged, motivated, and making continual progress \cite{plass2020toward}.

Learners also want to make sure that learning is \textbf{Aligned with Future Goals (G12)}.
\citet{knowles1973adult} says that adults are problem-oriented, and need to know how what they learn supports why they learn.
Students want to see how the topics they are learning relate to future careers. 
% (\textit{``create a, like a, a more, um, worldwide perspective of the career or the application of the topics we are learning''}).
AI tools should make clear how the content supports students' learning goals so adult learners stay engaged and motivated 
% (\textit{``engaging in learning when they see its relevance to their personal and professional lives''}) 
by giving students meaningful challenges that support learners' careers. 
By situating learning within contexts that matter to adults, AI tools can support persistence and engagement.
This guideline can enhance cognitive presence by connecting the content to authentic applications, and teaching presence by aligning the educational experience with the learner's long-term goals.

Learners stressed the importance of receiving relevant help when they need it. 
\textbf{Contextualized Help (G13)}  states that AI tools should recognize when learners need help and give them accurate, proactive instructions.
% (\textit{``system being able to identify when learners need help or guidance''}).
Students voiced frustration with the help methods currently available in AI tools,  
% ("\textit{I don't need more hints but I need to see how to move along when I am feeling stuck}")
while also raising concerns about trusting the accuracy of an AI's response 
(\textit{``the big [problem] is you're given a wrong answer, but you don't realize it''}).
AI tools must give accurate, contextualized answers without errors or hallucinations or user trust quickly erodes.
This guideline supports increasing cognitive presence by engaging with the student in the learning process through feedback loops, and teaching presence by providing clear instructional guidance.

\textbf{Actionable Insights (G14)} within AI technologies can increase teaching presence by helping instructors design instruction and increase cognitive presence by helping learners self-monitor their progress.
Teachers wanted to see who was using the tools, what activities they engaged in, 
% (\textit{``That'll be helpful too, knowing who accessed what and when.''})
and how the students progressed over time.
% (\textit{``It would be good to see how one student is doing in detail, and also how all the students are doing for different types of problems''}). 
Teachers then wanted to use this information to improve their instruction and content sequence. 
% (\textit{``If several students are getting hints or missing, I would go back to check what's going wrong and whether I need to re-teach that area''}). 
Students also voiced that data analytics help them keep pace in the class and reflect on their progress 
(\textit{``For me that helps me like just getting a visual representation of where I am and where I need to be and what I'm missing. That helps me a lot actually.''}). 
AI tools should move beyond presenting raw data to providing actionable insights that inform instructional decisions and support learner reflection \cite{yan2021including}.

\subsection{Social Presence}

Within the Community of Inquiry framework, social presence allows learners to project their authentic selves into the learning environment to interact with others in a meaningful way.
Students cited social connection with other humans as something they do not want to lose through interacting with AI tools. 
\textbf{Fostering community (G19)} should be a goal of AI tools, which could be realized by helping learners find peers with shared goals, interests, or contexts. 
Adult learners want to make friends, 
% (\textit{``I didn't find myself approached by anyone, even though I am someone who would like to make friends in the program.''})
not just for social connection, but also because learning communities are an important part of education. 
% (\textit{``A huge part of learning is, like, bouncing ideas off of each other.''}). 
Learning on your own can be disheartening, 
(\textit{``If I'm struggling with the class, it feels like I'm struggling alone.''})
but AI tools can help to offset those feelings both by providing content support, and connecting learners together.
% (\textit{``It's like a robot teacher because it's really, like, helping me, like, I didn't feel overwhelmed and all by myself.''}).
% (\textit{``I think a sense of community is, is really important''}).

\subsection{Teaching and Social Presence}

One of the most talked about concerns for adult learners was social connection. The following guidelines seek to increase social presence through lowering the barrier to social interaction and preserving interpersonal connections. They also can strengthen teaching presence by centering the instructor's responsibility for course design and intentionally creating ways for students to learn social skills.

Adult learners consistently emphasized the importance of human presence in their educational experiences, expressing greater trust in guidance from humans than from AI.
\textbf{Preserving the Human Touch (G17)} highlights how learners value feedback, explanations, and materials grounded in human judgment rather than automated responses.
Students worried that AI-driven instruction removes the interpersonal elements of learning, particularly when individual needs are involved.
% (\textit{``I think one thing that's true of AI all around, whether that's a tutor or not, is it takes out the human factor. The AI isn't gonna understand that there might be some barriers from it trying to walk you through something [like a learning disability] that it's not able to see.''}). 
Others described a broader concern that increasing reliance on technology makes education feel transactional and diminishes opportunities for meaningful connection with instructors.
While AI tools may increase access to information and efficiency, students cautioned that overdependence could erode patience, empathy, and tolerance for frustration 
(\textit{``I have the slippery slope idea that if people have these robot servants at their fingertips at all times, that they will have zero tolerance for frustration.''}).
% While adult learners are motivated by their future goals and value learning efficiency and task completion, they also emphasize the relationships that are fundamental to effective learning environments.
% AI tools that focus primarily on automation and efficiency may struggle to support the adult learner's relational needs.
To build trust and engagement, AI should supplement rather than replace human involvement to preserve the relational dimensions of learning that students find most meaningful.

Making genuine connections can be challenging, and soft skills like social interaction are skills that AI tools can help students learn. 
\textbf{Scaffolding Social Competencies (G18)} can help learners make friends while teaching them how to reach out to people without the assistance of an AI tool. 
Initiating conversations with strangers can be uncomfortable 
% (\textit{``It can be really daunting to like reach out to people, especially in like an online class with like hundreds of students.''})
, and asking for help in a large, public forum can feel intimidating. 
% (\textit{``I think one of the big barriers for me that's stopping me from talking to people is that [the online forum] is pretty public.''}). 
Consulting an AI tool with questions could lower the barrier to receiving help since students may feel that classmates or instructors will judge them. 
% (\textit{``Sometimes you think that you have a lot of dumb questions. And like, I have a lot of friends and I'll ask them, but I really just feel bad for asking. So if you're asking a robot, it might just feel like a lot better''}). 
At the same time, some students worried how over-reliance on AI may impact their real-world social skills 
(\textit{``If you over-rely on the tool, like, if you just purely use it as a source of meeting new people, it can kind of stand in the way of you meeting people in real life.''}). 
AI tools should not only connect students to build a learning community, but also support their development of social competencies so they will not need the tool to make future connections.

% think about if-then for issues? if you see this issue, do this thing? applicability of the design guidelines - maybe do this in the supplementary material with the cards. we have the stakeholder/problem, what other groupings can we include?

% broader sorting for the guidelines (eg. before, during interaction, over time, etc)

% checklist, link blue to pink to green

% discuss supplementary materials of a sankey chart linking blue to pink to green and the design cards

% present in a table? 
% Applying Design Guidelines to Improve AI Technologies
\section{Application of Design Guidelines}

\subsection{Heuristic Evaluation}
Heuristic evaluation lets researchers systematically analyze interactive systems in order to surface strengths, limitations, and design trade-offs in a structured and comparative manner \cite{nielsen1990heuristic}. Here, we operationalize our design guidelines by conducting a heuristic evaluation of the AI-ALOE technologies described in Section~\ref{section:ai_technologies}, using them as an analytic lens to examine how well these systems support adult learning needs and identifying opportunities for improvement. Because these technologies were developed without an explicit set of adult-learning-specific design guidelines, they provide a natural test case for illustrating how the guidelines can be used to support evaluation, ideation, and reflection.

\begin{table*}[]
\begin{tabular}{@{} l p{4cm} c p{7cm} @{}}
\toprule
\textbf{ID} & \textbf{Guideline}                                                               & \textbf{Sample Score} & \textbf{Rubric Description}\\ \midrule
G1          & [This tool] is open about data practices and share governance with learners.       & 3                     & Evaluate whether data collection, storage, use, and access are clearly communicated to users. Higher scores reflect meaningful user control (e.g., ability to review, manage, or delete data) and understandable disclosures. Lower scores reflect opacity, unclear retention or access, or limited user agency over personal information.\\
G2          & [This tool] is accessible and fit into the busy lives of instructors and learners. & 2                     & Evaluate whether the tool can be used quickly and reliably across contexts (work/home/school), times (including nights/weekends), and devices, with reasonable cost and effort. Higher scores reflect low barriers to access and flexible modes of engagement. Lower scores reflect restricted access contexts, or poor support for diverse needs.\\
G3          & [This tool] is informed by learning science and learning theories.           & 4                     & Evaluate whether the design rationale and core interactions are grounded in established learning theories. Higher scores reflect explicit theoretical grounding and design choices that apply theory to support comprehension, motivation, and knowledge construction. Lower scores reflect absent or inconsistent theoretical justification, or features that may conflict with relevant theory. \\ \bottomrule
\end{tabular}
\caption{Excerpt from the heuristic evaluation rubric, illustrating how AI tools could be evaluated against the guidelines for reflection and ideation.}
\Description{A table titled “Excerpt from the heuristic evaluation rubric” with four columns: ID, Guideline, Sample Score, and Rubric Description. It shows three example guidelines with corresponding scores and evaluation criteria:

G1: The tool is open about data practices and shares governance with learners; sample score of 3. The rubric evaluates whether data collection, storage, use, and access are clearly communicated. Higher scores indicate meaningful user control, such as the ability to review, manage, or delete data, and clear disclosures. Lower scores indicate lack of transparency, unclear data policies, or limited user control.
G2: The tool is accessible and fits into the busy lives of instructors and learners; sample score of 2. The rubric evaluates whether the tool can be used easily across contexts (such as home, work, or school), times (including nights and weekends), and devices, with reasonable cost and effort. Higher scores reflect flexible, low-barrier access, while lower scores reflect restricted access or poor support for diverse needs.
G3: The tool is informed by learning science and learning theories; sample score of 4. The rubric evaluates whether the design is grounded in established learning theories. Higher scores indicate clear theoretical foundations and design choices that support comprehension, motivation, and knowledge construction. Lower scores indicate weak or inconsistent theoretical grounding or potential conflicts with learning principles.

Overall, the table demonstrates how guidelines are paired with example scores and detailed criteria to support structured evaluation of AI learning tools.}
\label{tab:heuristic_checklist}
\end{table*}

\subsubsection{Technology Coding}
To carry out this evaluation, we recruited four independent coders to apply the full set of design guidelines to each of the AI-ALOE technologies. 
Coders worked independently and evaluated multiple systems, with each technology reviewed by two coders.
Coders relied on a combination of demo videos and peer-reviewed publications describing the technologies to understand user interactions and underlying design intentions.
Where aspects of a system were ambiguous or not immediately evident from demonstrations alone, coders consulted the associated literature to clarify system functionality.

Coders used a structured rubric to guide their evaluations. 
The rubric consisted of each guideline alongside its description, framing the guideline as an evaluative statement applied to a specific technology (e.g., “[This tool] is easy to understand and frictionless to use”). 
For each guideline, coders rated the degree to which here was evidence that the technology applied the guideline on a five-point Likert scale, ranging from strong evidence that the guideline was not met to strong evidence that it was satisfied. 
Coders were also required to provide a written justification explaining their assessment. 
An excerpt from the heuristic evaluation rubric is provided in Table~\ref{tab:heuristic_checklist}.

To resolve differences in ratings, coders participated in structured meetings to discuss their evaluation and how they applied each guideline. 
During these sessions, coders explained the reasoning behind their scores, clarified how they interpreted the intent and scope of the guidelines, and examined points of disagreement. 
This collaborative process allowed coders to refine their judgments and reach greater consistency in how guidelines were applied across technologies.

\begin{figure*}
    \centering
    \includegraphics[width=1\linewidth]{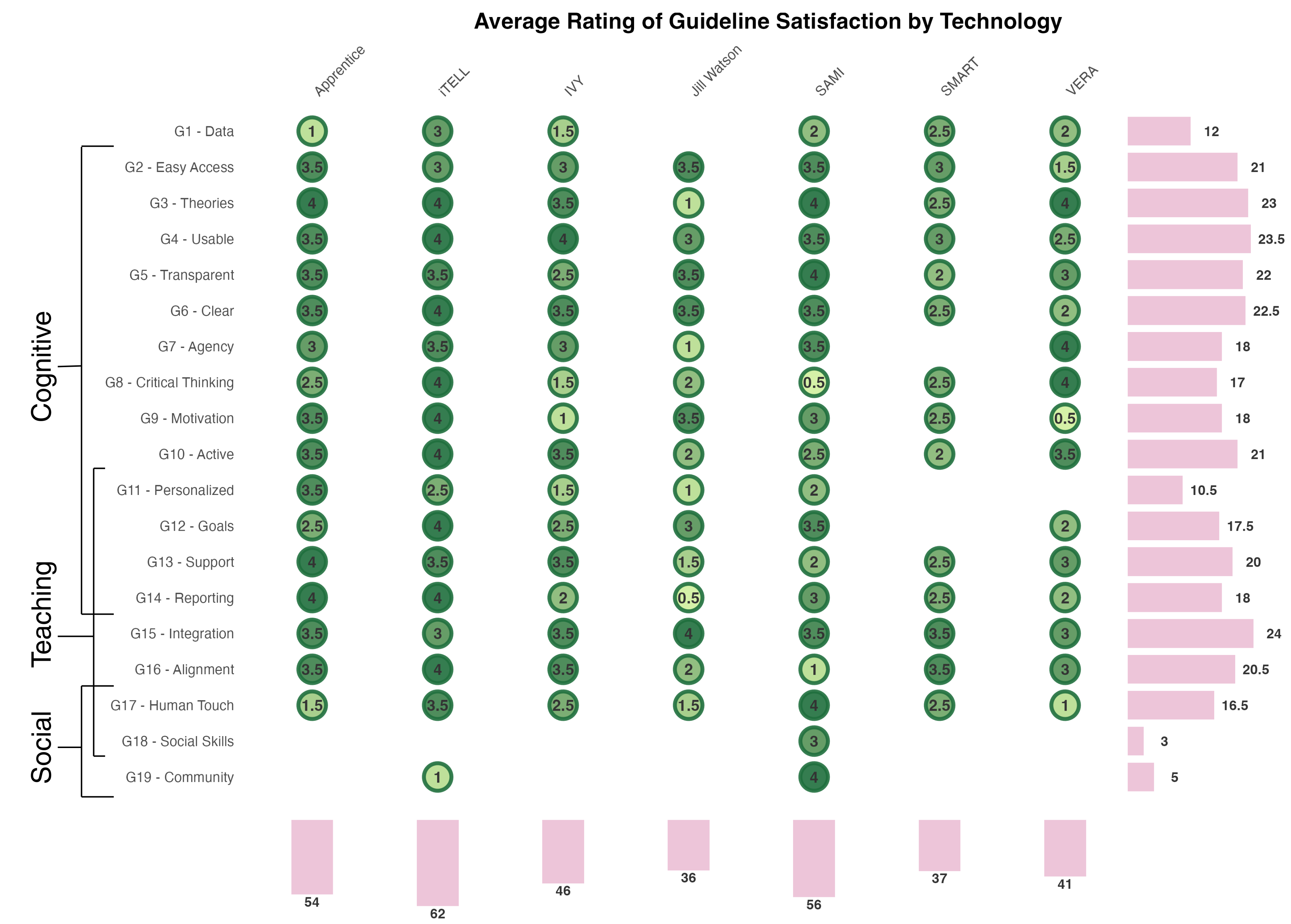}
    \caption{Average satisfaction ratings for each design guideline across the research institute's technologies.
    Green bubbles show average ratings by two coders on each technology from 0 (there is strong evidence the technology does not satisfy the guideline) to 4 (there is strong evidence the technology does satisfy the guideline).
    If the guideline was rated as 0 by both coders, the bubble is missing to highlight a potential improvement area for the technology.
    Pink bars on the bottom show the total rating across all technologies, and the pink bars on the right show the total rating across each guideline. 
    Guidelines vary in satisfaction, with the strongest support for integration, usability, learning theories, cognitive load management, and transparency  (G15, G4, G3, G6, G5), and the lowest support for social skills, community, personalization, and open data practices (G18, G19, G11, G1).}
    \label{fig:satisfaction}
    \Description{
    The figure presents a bubble chart showing coder ratings (0-4 scale) of seven technologies against 19 design guidelines (G1-G19). 
    Each bubble represents the average score for a guideline-technology pair, with darker green bubbles indicating higher satisfaction and lighter green/yellow bubbles indicating lower satisfaction. 
    Scores closer to 4 mean strong agreement that the technology satisfied the guideline. 
    Scores closer to 0 mean strong agreement that the technology did not satisfy the guideline.
    A score of 0 is represented as an empty space in the graph.

    The bar chart on the right summarizes total satisfaction across all technologies for each guideline. 
    Guidelines with consistently high scores include G15 (integration with other tools), G4 (usability), and G3 (informed by learning science). 
    Lower-scoring guidelines include G19 (fostering community), G18 (scaffolding social competencies), and G11 (personalization).
    
    Along the bottom, horizontal pink bars show the total satisfaction per technology across all guidelines. 
    Technologies with the highest scores are iTELL, SAMI, and Apprentice. 
    Technologies with lower scores are Jill Watson and SMART.}
\end{figure*}

After consensus was reached, we computed the average rating for each technology-guideline pair and shared these results with the research and development teams responsible for each system. 
Teams were invited to review the evaluations and provide additional documentation if they believed aspects of their technology had been mischaracterized. 
% Only one team elected to submit a response. 
Coders reviewed the additional materials, revised ratings where appropriate, and reconvened to discuss any changes.
This process resulted in a strong overall agreement across coders (intra-class correlation coefficient = 0.80) when determining whether a guideline was satisfied. 
The aggregated results of this evaluation are summarized in Figure~\ref{fig:satisfaction}.

\subsubsection{Results}
Across the evaluated technologies, several guidelines were consistently rated as satisfied. 
The highest average scores were associated with usability  (\textbf{G4}) and integration with existing educational technology ecosystems (\textbf{G15}). 
Most tools were compliant with the Learning Tools Interoperability (LTI) standard and did not require substantial additional setup or configuration when used. 
The technologies also use familiar web-based interfaces, like chat interfaces, textboxes, and drag-and-drop components, to allow users to onboard themselves to the technology quickly.

A second cluster of highly rated guidelines reflected cognitive and instructional design theories: \textbf{G3} (informed by learning science), \textbf{G6} (minimizing cognitive load), and \textbf{G10} (active, constructive, and interactive engagement). 
Interfaces were designed to reduce distractions and focus attention on what matters most.
Learning activities frequently required active participation, exploration, and reflection, rather than passive content consumption. 
As a result, the evaluated tools were consistently rated as supporting interactions grounded in well-established instructional principles.

The technologies also scored highly on \textbf{G5} (transparent and explainable). 
In particular, systems often provided explicit explanations for feedback and guidance, especially in response to incorrect learner actions. 
These explanations helped clarify why particular forms of feedback were presented. 
Across technologies, this form of explainability was more prevalent in learner-facing feedback mechanisms than in system-level descriptions of underlying models or processes.

In contrast, several guidelines were consistently rated as less well supported. 
The lowest average scores were associated with social interaction and community-related guidelines, including \textbf{G18} (scaffolding social competencies) and \textbf{G19} (fostering community). 
For most technologies, opportunities for peer-to-peer interaction were limited, and explicit support for developing social skills was either minimal or absent. 
These patterns were consistent across tools, with few systems incorporating features explicitly designed to support sustained social engagement.

Another guideline that received low satisfaction scores across technologies was \textbf{G1}, which concerns transparency around data practices. 
Coders noted that information about data collection, storage, access, and use was often not readily visible to users through system interfaces. 
As a result, it was difficult to determine how user data were being handled or what degree of control users had over their information.

\textbf{G11} (personalization) was another area with relatively low satisfaction. 
While many technologies incorporated surface-level personalization -- such as adapting content examples, responding to learner inputs, or providing regular knowledge checks -- fewer systems used learner data for adaptations such as  task sequencing or calibrating difficulty. 
Personalization was therefore present in limited forms but was not consistently reflected in deeper structural adaptation across learning activities.

Satisfaction scores varied across individual technologies, revealing distinct strengths and limitations. 
For example, SAMI, as a social agent designed to connect learners,  was the only system that consistently satisfied the social presence guidelines (\textbf{G18}, \textbf{G19}). 
At the same time, SAMI received lower scores on guidelines related to metacognitive support (\textbf{G8}), personalization (\textbf{G11}), and instructional alignment (\textbf{G16}) since these fall outside its primary function. 
Jill Watson, as a conversational virtual TA, engages and empowers students to ask questions and quiz themselves, building motivation and self-efficacy, so it showed strong support for learner motivation and engagement (\textbf{G9}), while receiving lower ratings for grounding in learning science theories (\textbf{G3}). 
VERA's ability to build scientific concepts through experimentation and scientific investigation earned a high rating on learning science-related guidelines (\textbf{G3}, \textbf{G7}, \textbf{G8}), but the tool received lower scored on motivation-oriented guidelines (\textbf{G9}, \textbf{G11}).

Overall, no single technology satisfied all 19 guidelines. 
Instead, the heuristic evaluation revealed variation in guideline coverage that aligned with the functional focus of each system. 
Technologies designed to prioritize particular instructional, cognitive, or engagement-related goals tended to support corresponding guidelines more strongly, and providing minimal and limited support for others.
When considered collectively, however, the broader ecosystem of AI-ALOE technologies provided coverage across the full set of guidelines. 
From this perspective, lower satisfaction scores on individual guidelines reflect trade-offs in scoped design decisions rather than an absence of support at the ecosystem level.

% The aggregated results of these evaluations are shown in Figure~\ref{fig:satisfaction}, which summarizes average satisfaction scores across guidelines and technologies.

% although a technology that addresses as many of these as possible would support learners in more diverse ways, we acknowledge some technologies will be targeted toward specific functions that (hopefully) fit in a broader ecosystem and not all guidelines will apply given the function the technology is taking on

% for guidelines that do apply to a technology, there can be several ways to improve/issues to address under that dimension (and then reference the underlying pink labels as an example). I.e. there’s generally an opportunity for improvement in some regard

\subsection{Guideline Explorer}\label{sec:guideline_explorer}

\begin{figure*}
    \centering
    \includegraphics[width=1\linewidth]{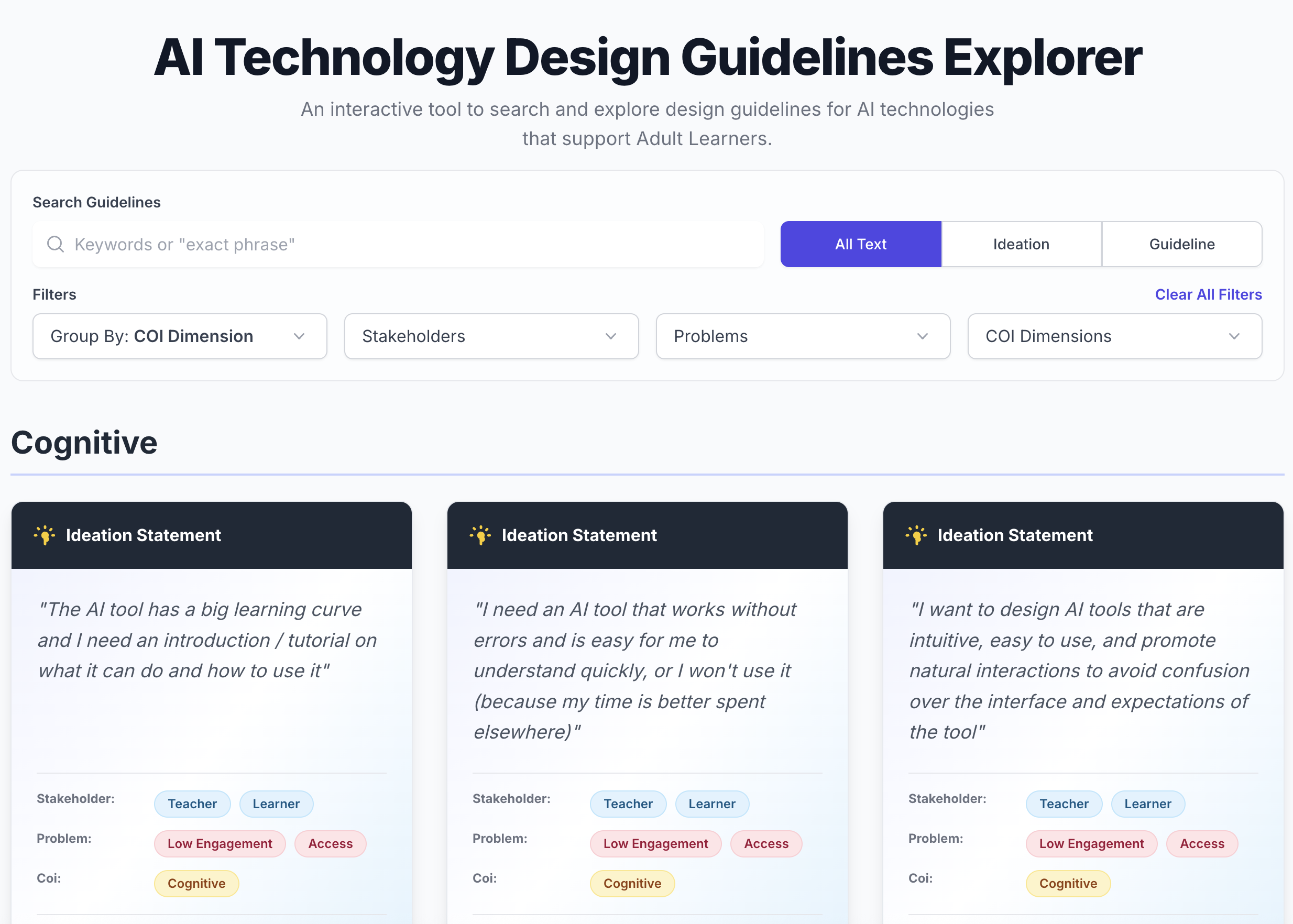}
    \caption{The Guideline Explorer interface for browsing and filtering ideation statements associated with each design guideline. The explorer allows designers to search statements and filter them by guideline, stakeholder, Community of Inquiry (CoI) dimension, and problem type to support stakeholder-centered ideation.}
    \label{fig:guideline_explorer}
    \Description{
    The figure shows the interface of the AI Technology Design Guidelines Explorer, an interactive tool for browsing and filtering ideation statements related to the design guidelines. 
    At the top of the interface is a search bar that allows users to enter keywords or exact phrases. 
    Below the search bar are filter controls that enable grouping and filtering by Community of Inquiry (CoI) dimension, stakeholder, problem type, and guideline.
    
    The main content area displays a section labeled “Cognitive,” indicating that the current view is grouped by CoI dimension. 
    Beneath this heading are multiple cards, each representing an ideation statement phrased as a first-person request from a user. 
    For example, statements express needs such as wanting an introduction or tutorial, requiring tools that are easy to understand, or desiring intuitive and natural interactions.
    
    Each ideation card includes tags that identify the relevant stakeholder (such as teacher or learner), the associated problem category (for example, low engagement or access), and the CoI dimension (in this case, cognitive). 
    }
    
\end{figure*}

We developed the Guideline Explorer (Figure~\ref{fig:guideline_explorer}) as an interactive tool that provides direct access to the ideation statements---the first-person expressions derived from the blue sticky notes generated during the affinity diagramming process described in Section~\ref{section:thematic_analysis}. These statements capture recurring needs articulated by learners, instructors, administrators, and developers across deployments. To make the data navigable, we annotated each ideation statement with metadata including its associated guideline, Community of Inquiry dimension, affected stakeholder group, and problem type. The Guideline Explorer enables designers to browse, search, and filter ideation statements along these dimensions. We envision designers first using the heuristic evaluation rubric to identify opportunities for growth within a technology, and then turning to the explorer to examine associated stakeholder needs.

A contribution of the explorer is that it makes qualitative design research directly accessible and actionable beyond our specific research context. When designing or refining AI-powered education technologies, developers can read stakeholder statements from actual adult learning contexts to help determine if and how they might incorporate specific guidelines in their technology. By engaging with ideation statements rather than specific, prescriptive solutions, designers can brainstorm and evaluate potential responses that are sensitive to their own context. We should note that although some of the ideation statements explicitly mention AI, one should not assume that AI-based solutions are always appropriate. Instead, it surfaces stakeholder concerns that may be addressed through a range of means.

Next, we present two cases detailing how the explorer might be used.

\subsubsection{Case 1}\label{sec:case_study_1}
Imagine a team used the heuristic evaluation and found that its technology was rated low on guideline \textbf{G14} -- \textit{AI tools should provide actionable insights to instructors, learners, and researchers}.
The research team knows that they are collecting lots of data, but the main use of that data is learning engineering and their own analyses.
The AI tool provides actionable insights to researchers, but not learners or instructors.
They open up the Explorer and investigate the \textbf{G14} ideation statements:

\begin{itemize}
    \item \textbf{Instructors and Researchers:} I want to know who are using the tools, who did what, and to track their progress.
    \item \textbf{Instructors and Students:} I want detailed learning analytics (through visualizations) so I can make better instructional decisions.
    \item \textbf{Researchers:} I want the tools to collect data and use it to facilitate teacher and tool personalization.
    \item \textbf{Researchers:} Learning tools should be designed for continuous improvements and learning engineering through historical data.
\end{itemize}

The team can now brainstorm ways to meet the stakeholder needs around actionable insights. 
Perhaps the data is not yet in a consumable format for stakeholders outside the research team, so they focus on transforming the data into something more interpretable like a table that provides an overview of class actions to meet the first ideation statement. 
Or, they could brainstorm ways to personalize the tool based on data so that each user has an experience tailored to their needs, meeting the third ideation statement. 
Maybe the team decides to build an instructor dashboard to meet the first statement, detailing who, what, and when, or a more detailed visualization of class-level misconceptions to help instructors structure in-class time around student needs.
Because self-directed adult students also play a role in instructional decision making, perhaps the team proposes a student-facing dashboard to make student progress visible and help students monitor and reflect on their learning trajectory.

\begin{figure*}
    \centering
    \includegraphics[width=1\linewidth]{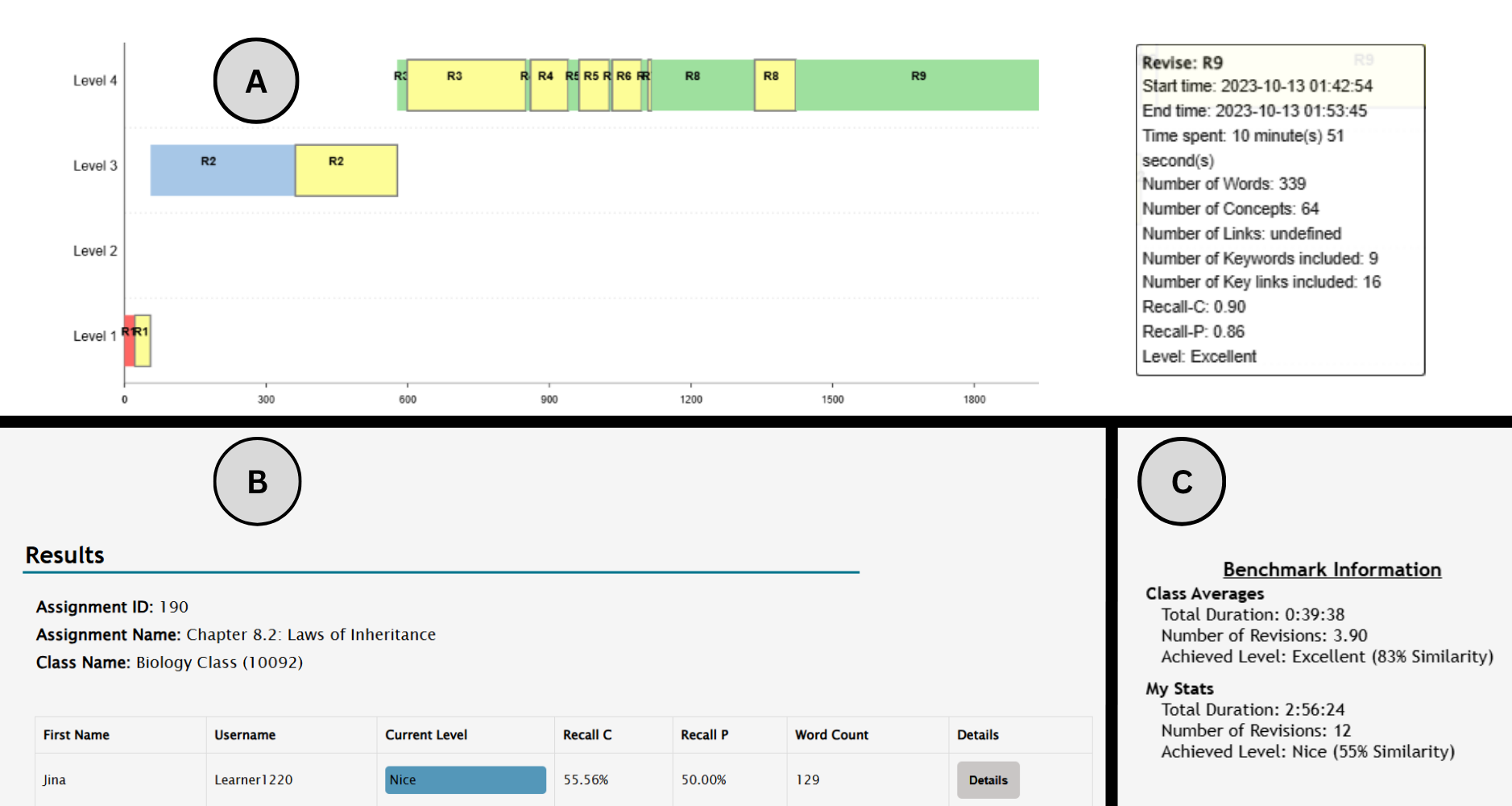}
    \caption{Example interfaces from SMART illustrating how learner data can be transformed into actionable insights for different stakeholders, supporting Guideline G14. Three panels show new changes made in response to learner and instructor needs. Panel A shows a timeline-based visualization of learner activity and revisions. Colored blocks represent different revisions or learning events over time, and hovering over a block reveals detailed metrics such as time spent, number of words, and number of concepts for fine-grained analysis of learning behaviors. Panel B presents an instructor-facing table summarizing student performance and progress. This view supports instructors in monitoring class progress, identifying students who may need support, and making instructional decisions based on summarized learning analytics. Panel C displays benchmark and individual statistics for reflection and comparison. Metrics such as total duration, number of revisions, and achieved level are shown for both the class average and the individual student, supporting reflection and self-monitoring.}
    \label{fig:SMART_Dashboards}
    \Description{
    The figure contains three panels labeled A, B, and C, showing example dashboards from SMART that illustrate how learner data can be presented as actionable insights for multiple stakeholders.
    
    Panel A displays a timeline visualization of a learner’s activity across multiple levels of an assignment. 
    Colored blocks represent different revisions or learning events over time, allowing users to see when changes were made, how long was spent, and how performance evolved. 
    Hovering over a block reveals detailed metrics such as time spent, number of words, number of concepts, and recall scores, for fine-grained analysis of learning behaviors.
    
    Panel B shows an instructor-facing results table. 
    Each row corresponds to a student and includes fields such as current level, recall measures, word count, and a button to view additional details. 
    This view supports instructors in monitoring class progress, identifying students who may need support, and making instructional decisions based on summarized learning analytics.
    
    Panel C presents benchmark information comparing class-level statistics with individual learner performance. 
    Metrics such as total duration, number of revisions, and achieved level are shown for both the class average and the individual student, supporting reflection and self-monitoring.
    }
\end{figure*}

Or, like SMART, the team could implement all of these ideas (Figure~\ref{fig:SMART_Dashboards}). 
After reviewing the user interview transcripts, the SMART team came to similar conclusions about how to improve their tool. 
If resources like the guidelines and the explorer had been available earlier, the team could have used them to more quickly identify promising areas for improvement. 
This does not suggest that teams should not engage with users; rather, it demonstrates how these resources can efficiently synthesize prior research into actionable design directions.

\subsubsection{Case 2}
Alternatively, assume a technology team is struggling with low adoption.
They developed their technology with the input of users and stakeholders from one university, but struggle trying to move their technology to another.
Their technology is embedded in several classes and they see many initial launches, but very few users continue to interact with the tool beyond the first few clicks.
The team decides to investigate stakeholder concerns related to low adoption using the explorer.
They find the following ideation statements:

\begin{itemize}
    \item \textbf{Instructors:} I won't use the AI if it isn't aligned with my teaching approach and style.
    \item \textbf{Instructors and Students:} I want ways to appropriately integrate AI with my other tools and to effectively manage AI tool notifications.
    \item \textbf{Instructors and Students:} I want options/choices for how I engage with content, as well as access to supplementary content.
    \item \textbf{Students and Researchers:} I want AI to provide explanations and sources from human experts so I can believe what the AI says is correct.
\end{itemize}

The team can now brainstorm ways to improve their tool toward greater adoption. 
The first point is a statement from instructors, so the team could investigate the instructional practices of the new school.
How well does their tool align with and reinforce the existing instructional practices (\textbf{G16})?
The second point is from both instructors and students about integrating into existing technology ecosystems (\textbf{G15}).
Is the AI tool connected to the other tools currently in use?
Are the notifications from all tools easily consolidated?
Can students and instructors efficiently navigate between each technology?
The third point is a concern raised by students asking for agency over their learning (\textbf{G7}).
Does the tool provide options for students or is there only one way to engage?
Are students locked into a prescribed path or can they make meaningful choices about their learning trajectory?
Does the tool provide optional, supplementary content and resources?
The forth point is also a concern from students regarding transparency and explanations within AI tools (\textbf{G5}).
Does the AI tool link back to human-crafted source material to justify its explanations?
Does the AI tool build trust by proving that its content is correct?

These questions show how the explorer can also support designers in looking across guidelines to uncover potential root causes of design challenges such as low adoption. 
Instead of treating a lack of adoption as a single problem, the team can trace it to interacting issues of instructional alignment (\textbf{G16}), ecosystem integration (\textbf{G15}), learner agency (\textbf{G7}), and transparency and trust (\textbf{G5}). 
Now the design team can make concrete changes to their tool, such as modifying instructional workflows, enabling integration with existing platforms, offering multiple learning paths, or adding citations and explanations to AI responses.
This kind of synthesis exposes relationships among guidelines that would be difficult to see when they are considered independently.
The explorer can help designers move from vague problem statements to coherent hypotheses about why a tool is not working through cross-guideline analysis.

\section{Discussion}

The guidelines presented in this paper are grounded in the perspectives, concerns, and desires of key stakeholders: learners, instructors, and researchers.
They reflect priorities from participants’ lived experiences with AI-enhanced learning technologies.
Even when individual guidelines align with principles that are already discussed in prior literature, empirically demonstrating their significance for this specific population of adult learners constitutes a meaningful contribution.
Many of the principles, like transparency, accessibility, and cognitive load management, are rooted in foundational learning science principles and are applicable across age groups and educational settings.
Their inclusion here demonstrates their relevance in practice with this under-studied population.
Validating these principles in the context of adult learning confirms their importance while grounding them in the realities of this population.

At the same time, the guidelines surfaced many concerns that are specific to the adult learner.
Adult learners have many responsibilities competing for their time, so they must divide their attention.
AI educational tools should be easy to access wherever, whenever (\textbf{G2}), be easy to use and understand (\textbf{G4}), and designed to minimize extraneous cognitive load (\textbf{G6}) so that learning can fit into adults' limited and fragmented time.
Adults are also more likely to be concerned with data governance (\textbf{G1}) than child learners. 
In addition to minimal time, adults are career-oriented, so learning experiences should be purposeful and connected to the learner's real-world goals (\textbf{G12}).
Adults have varied backgrounds, and bring a wealth of prior knowledge and experiences into the classroom.
AI tools should provide personalized pathways based on the adult learner's background (\textbf{G11}), provide timely and contextualized support when needed (\textbf{G13}), and give learners meaningful agency over how they engage with and direct their learning (\textbf{G7}). 
These guidelines operationalize broader design principles in ways that are directly responsive to the needs of adult learners.

We view these adult-focused guidelines through the lens of universal design: design choices that address the needs of adult learners often improve the learning experience for all users. 
Universal Design \cite{CEUD_2026} is a framework originally developed for the built environment, but these principles have since been applied broadly to communication, technology, and education, based on the insight that design choices that address the needs of users with diverse abilities, constraints, or preferences often improve the experience for everyone.
Universal Design emphasizes flexibility in use, guiding designers to ``Provide choice in methods of use'' like our \textbf{G7}, giving learners agency over their learning, and ``Provide adaptability to the user's pace'' like our \textbf{G11}, where AI tools should provide individualized pathways for learning based on learners' knowledge, skills, and abilities.
The framework also emphasizes presenting information in easy-to-understand, accessible ways (``Provide compatibility with a variety of devices.'' and ``Eliminate unnecessary complexity.'') similar to \textbf{G2} and \textbf{G6}, which strive to make AI tools accessible whenever, wherever and present information clearly and simply.

Building on this foundation, Universal Design for Learning (UDL) applies these principles specifically to educational contexts through three core principles: multiple means of representation (how information is presented), action and expression (how learners demonstrate knowledge), and engagement (how learners are motivated and supported) \cite{cast_2024}.
Features such as flexible access and personalized support address representation by ensuring information reaches learners in forms they can perceive and process. 
Learner choice, agency, and goal-oriented experiences support action and expression by giving learners control over how they demonstrate and apply knowledge. 
And motivational scaffolds, social connection, and metacognitive development exemplify UDL's engagement principle, which emphasizes sustaining effort, welcoming identities, and cultivating emotional capacity.
% Features such as flexible access, personalized support, learner choice and agency, and goal-oriented educational experiences may be especially important for adults, but they also benefit all learners with limited time, varied backgrounds, or different learning preferences across educational contexts. 
Examining the unique needs of adults surfaces requirements around flexibility, relevance, and efficiency that would benefit all learners with limited time, varied backgrounds, or different learning preferences across educational contexts.
In this sense, focusing on the constraints and priorities of adult learners does not narrow the applicability of these guidelines.
Rather, it strengthens their general value by promoting more inclusive, usable, and effective AI-enabled learning technologies for a broad range of learners.

% ethical considerations
One issue that came up during our discussions with AI-ALOE researchers was whether there should be an additional guideline concerning the ethical use of AI.
Although the guidelines do not explicitly frame themselves as ethical principles, we believe several provide concrete and actionable ways to build AI technologies with ethics in mind.
% The ethical concerns raised by stakeholders are encapsulated within the specific design practices described in this paper.
For example, transparency and explanations (\textbf{G5}) is an actionable response to concerns about hallucinations, misinformation, and other limitations of AI-generated content.
Data governance (\textbf{G1}) translates into making user privacy and consent visible and giving users more ownership of their data.
Guidelines around learner agency (\textbf{G7}), contextualized support (\textbf{G13}), instructional alignment (\textbf{G16}), and preserving the social elements of learning (\textbf{G17}-\textbf{G19}) reflect ethical considerations around autonomy, accountability, and the risk of losing human connections.
The guidelines offer an actionable path for embedding ethical considerations into AI-supported learning environments.

% unequal distribution of guidelines
An important pattern that emerged from our analysis is the unequal distribution of design guidelines and empirical support across the Community of Inquiry dimensions. 
While there is a strong emphasis on cognitive and teaching presence, there is comparatively little attention to social presence.
Of the 19 guidelines, only three (\textbf{G17}-\textbf{G19}) explicitly address the social dimensions of learning, and these guidelines received the lowest satisfaction ratings across the AI-ALOE technologies.
Even when AI-ALOE tools included features to support social presence, they were often peripheral rather than a central function, with SAMI being a notable exception.
This imbalance suggests that AI-supported adult learning tools are relatively mature in supporting individual cognition, usability, and instructional alignment, but underdeveloped in fostering social connection, building community, and developing interpersonal skills.

In addition, concerns about social presence and the erosion of human elements in online learning environments were primarily expressed by students (see Figure~\ref{fig:stakeholder_comments_distribution}).
Learners repeatedly emphasized the importance of interpersonal interaction and human judgment in learning, apprehensive that AI systems could reduce opportunities for meaningful connection with instructors and peers.
Participants also worried that AI tutors and automated feedback could hide contextual factors such as learning disabilities, emotional states, or personal struggles that humans are better equipped to recognize.
Learners also expressed concerns about blindly accepting information from an AI; e.g., in case the virtual TA reports untrue information about assignment deadlines or an AI tutor provides incorrect remediation.

Adult learners have clearly articulated that learning is not solely a cognitive activity.
Relationships, community, and connection are essential components of a learning environment.
Students continue to prefer learning from human experts, as AI systems have not yet demonstrated the trustworthiness, accountability, contextual understanding, and human presence required to be accepted as instructional authorities.
Future AI learning technologies should move beyond performance-oriented goals and more intentionally design for social presence and shared inquiry.

% tensions within the guidelines
% Our analysis also reveals potential tensions within the guidelines that designers must navigate. 
In addition to gaps in social presence, our analysis reveals potential tensions within the guidelines that reflect differing priorities across stakeholder groups. 
One such tension exists between the guidelines that AI tools should be easy to understand and frictionless to use (\textbf{G4}) while providing quick and proactive help (\textbf{G13}), and the guidelines that emphasize meaningful challenge (\textbf{G8}), metacognitive development (\textbf{G10}), and active engagement (\textbf{G12}). 
Learners and instructors wanted to reduce usability friction, like confusing interfaces, unclear expectations, or excessive onboarding demands, as it makes the tool feel efficient and supports ongoing work.
Delays or unclear responses were frustrating to users, especially if students wanted to get the answer and move on.
% (\textit{``The tutor does not give instant feedback for the student, if they are feeling stuck and want to move along''}).
% However, frictionless design should not be interpreted as eliminating difficulty from learning itself, but rather as removing extraneous barriers to distract from effort and persistence.
However, the desire for immediacy may conflict with opportunities for learners to engage in deeper thinking.
Quickly providing answers can reduce the time and cognitive effort students spend generating and evaluating their own responses, potentially limiting reflection and understanding.
Rather than optimizing for speed or responsiveness, AI tools may need to balance timely support with mechanisms that encourage learners to pause, consider their own reasoning, and remain cognitively engaged.
% Prior work suggests that not all friction is detrimental. 
When intentionally designed, friction can promote slower, more mindful engagement and improve users’ understanding of task goals and overall satisfaction \cite{mejtoft2019design}.
Well-designed AI tools can preserve productive struggle through meaningful challenges and reflective prompts while minimizing operational friction that distracts from learning goals.

At the same time, researchers were the only stakeholder group to emphasize that AI learning technologies should be grounded in learning theory and empirical evidence of effectiveness.
The competing desires between students, instructors, and researchers suggests a risk that systems optimized for immediacy and ease of use may be preferred by learners and instructors, even when those designs may undermine learning processes that rely on productive struggle and sustained cognitive effort. 
When ease of use is a significant deciding factor for adoption, prioritizing convenience over reflection and sense-making becomes a danger for learning technologies.
Future AI learning technologies should make the value of productive struggle more visible to users, so they better understand the benefits of slowing down and engaging in deliberate practice.

% lag time - contextualized help 
% JillWatson-Learners Focus Group-39: usually when you get a question in your head, it's not, uh, pre-planned, and you, you hope that you can get the answer as soon as possible so that you can, um, move on with whatever work you're doing.

% It's frustrating when you ask it something and it doesn't provide the answer

% The tutor does not give instant feedback for the student, if they are feeling stuck and want to move along

% self-effacacy/motivation - frustration
% For my wellbeing, I was like... pfffff... this is a waste of time and I was frustrated and I didn't see any point

% I didn’t know what it wanted me to enter.

Another tension arises between learner agency (\textbf{G7}), personalization (\textbf{G11}) and instructional alignment (\textbf{G16}).
Adult learning theory emphasizes autonomy, self-direction, and choice, and learners frequently expressed a desire for control over pacing, support, and learning pathways.
AI systems are well-suited to providing many options for students through personalization, optional scaffolds, and adaptive pathways.
However, instructors voiced parallel concerns about alignment with course goals and assessments.
Learner agency in adult education does not imply total freedom, but rather meaningful choice within a structured instructional context.
When AI tools make instructional goals, expectations, and rationales visible, they enable learners to exercise agency in relation to those goals.
Instructional alignment makes agency productive by orienting learner's choices around pacing, strategy, and support toward course objectives.
% Alignment can also strengthen learner agency by increasing learner adoption and engagement with the course by explaining how choices fit into the broader instructional goals.
AI tools can translate instructor intent to students in a way that preserves meaningful learner autonomy.

The guidelines have already informed concrete revisions across multiple systems, including Apprentice, iTELL, Jill Watson and SMART.
% demonstrates practical value
Jill Watson, iTELL, and Apprentice were revised to make all features compatible on mobile devices, making the tool accessible from many locations and times (\textbf{G2}).
The iTELL development team enabled universal integration with industry learning management systems using SCORM broker files (\textbf{G15}), and added AI-generated feedback for constructed responses to ensure that support is contextualized and relevant (\textbf{G13}).
Apprentice also expanded its feedback mechanisms to offer deeper hints, supplementary videos and texts, and problem-specific tutorials (\textbf{G13}).
In addition, Apprentice developers collaborated with instructors to create multiple versions of the same tutored content to ensure alignment with instructional practices (\textbf{G16}).
SMART also incorporated new dashboard features to learners and instructors, as described in Section~\ref{sec:case_study_1}.
These examples demonstrate how stakeholder-informed guidelines can be translated into concrete features across platforms.

% not all technologies need to satisfy all guidelines, ecosystem perspective

% EXPAND ON THE ECOSYSTEM PERSPECTIVE

Our heuristic evaluation revealed significant variation in how individual technologies align with the design guidelines.
Systems tended to strongly support guidelines that were closely aligned with their primary instructional purpose and functional scope.
This variation revealed clear trade-offs in design focus.
Technologies that emphasized depth in a particular dimension of adult learning often did so by deprioritizing features associated with other dimensions.
As a result, no individual system demonstrated strong support across all 19 guidelines.
When considered collectively, however, the AI-ALOE technologies provided strong coverage across the full set of guidelines.

These results suggest that the guidelines can support coordination across an ecosystem of complementary technologies, rather than requirements for any single system.
Using the Community of Inquiry framing, AI technologies may choose to prioritize cognitive, teaching, or social presence, and focus on the set of guidelines that align with the goals of the tool. while leaving others unsatisfied. 
However, there may be some guidelines that are 'non-negotiable` to users and may impact the tools' usefulness and broader adoption. 
Guidelines around usability (\textbf{G4}), accessibility (\textbf{G2}), and instructional alignment (\textbf{G15}) may need to be satisfied by every tool within an educational ecosystem. 
A tool that is confusing, inaccessible, and does not support instruction undermines the learning experience, and no other tool can compensate for that failure.
Other guidelines like \textbf{G11} and \textbf{G19} could be distributed across a set of complementary tools, allowing for a deeper personalization as learners navigate the ecosystem.
Building a singular tool that satisfies all 19 guidelines would be a difficult task, but a thoughtfully coordinated ecosystem can achieve what no single tool can alone.
In this framing, the guidelines can help designers, instructors, and institutions reason about priorities, trade-offs, and how multiple tools can work together to support the diverse cognitive, teaching, and social needs of adult learners.

\section{Limitations and Future Work}

Our guidelines should be interpreted with respect to the context from which we collected data, that is, adult learning environments. 
% Our analysis focused on a set of AI technologies developed within a single research institute.
While these tools represent a wide variety of functions, they may not capture the full spectrum of educational technologies in use across different institutions and contexts.
Though we have demonstrated the practicality and applicability of the guidelines to produce meaningful change in AI technologies for adult learners, it is unclear if and how the guidelines can be applied beyond adult and online learning contexts.
Some of the guidelines may generalize across different age groups and modalities, while others, such as scaffolding social competencies and supporting career-oriented goals, may be specific to adult learners balancing professional, personal, and educational responsibilities.

In addition, our analysis relied on thematic coding and heuristic evaluation.
% from a small set of researchers and coders. 
Although we used reflexive thematic analysis and consensus-building sessions to strengthen reliability, the judgments necessarily reflect researcher and coder interpretation.
While the guidelines capture stakeholder concerns and priorities, further empirical validation is needed to assess their impact on learning, engagement, and adoption in practice.

Future work should extend these guidelines through empirical validation, examining how these guidelines relate to learning outcomes, motivation, and technology use in authentic instructional settings.
Cross-institutional studies, such as workforce training, community colleges, or international contexts, could test whether the guidelines hold in settings with different learner demographics, institutional structures, and resource constraints.
Such work would help determine which guidelines are broadly applicable and which are more context-specific.

\section{Conclusion}
Through a reflexive thematic analysis of stakeholder feedback, we derived 19 design guidelines that articulate recurring needs, challenges, and values across adult learners, instructors, and researchers. 
Organized through the Community of Inquiry framework, these guidelines emphasize the cognitive, teaching, and social dimensions of adult learning while also addressing foundational concerns such as transparency, data governance, and accessibility. 
They provide a structured vocabulary for reasoning about AI-supported learning beyond technical performance, highlighting issues of agency, trust, alignment with instructional practice, and the preservation of human connection.

We demonstrated the practical utility of these guidelines through a heuristic evaluation of the AI-ALOE technologies. 
This evaluation revealed both strengths, such as usability, theoretical grounding, and explainability, and areas for growth, including support for social connection, personalization, and open data practices. 
% Importantly, this analysis revealed that no single system can or should satisfy every guideline. 
Rather than functioning as a checklist, the guidelines are intended to support reflective design: making trade-offs visible, clarifying whose needs are prioritized, and enabling teams to reason about how individual tools contribute to a broader learning ecosystem.

To support this process, we introduced the Guideline Explorer as a companion that links each guideline to underlying stakeholder concerns and ideation statements. 
The Explorer enables designers to move from abstract principles to concrete design opportunities by making the guidelines and supporting affinity diagram easily navigable and searchable. 
The guidelines, heuristic rubric, and Explorer offer an approach for evaluating existing systems, informing new designs, and facilitating discussion among interdisciplinary teams.

The guidelines are derived from technologies developed within the AI-ALOE research institute and reflect the contexts, populations, and design priorities of those deployments. 
While the setting spans diverse disciplines and institutions, further research is needed to examine how the guidelines generalize to other adult learning environments, including workplace training, informal learning communities, and industry-developed platforms. 
As generative AI and adaptive systems become increasingly prevalent in education, there is a growing need for empirically grounded frameworks that help align technological innovation with the realities of adult learning.
AI is not the solution to all educational technology problems, but analyzing the concerns voiced by users can ground ideation and problem-solving in stakeholder needs.

Ultimately, our work aims to support lifelong learning by supporting the design of AI technologies that are grounded in adult learners’ needs, resources, challenges, and lived contexts.
Designing for adult education requires more than adapting tools built for younger learners.
It demands sustained attention to the social, cognitive, and institutional conditions in which adults learn.
By designing learning tools based on adult learners' needs, resources, challenges, and lifestyles, researchers and technology designers can improve engagement, social connections, and learning outcomes for adult learners specifically.
The guidelines presented in this paper lay the foundation for designing AI that respects and supports the complex lives of adult learners.

%%
%% The acknowledgments section is defined using the ``acks'' environment
%% (and NOT an unnumbered section). This ensures the proper
%% identification of the section in the article metadata, and the
%% consistent spelling of the heading.
\begin{acks}
The material in this paper is based upon work supported by the National Science Foundation under Grant No. 2247790 and Grant No. 2112532. Any opinions, findings, and conclusions or recommendations expressed in this material are those of the author(s) and do not necessarily reflect the views of the National Science Foundation.
\end{acks}

%%
%% The next two lines define the bibliography style to be used, and
%% the bibliography file.
\bibliographystyle{ACM-Reference-Format}
\bibliography{paper}

%%
%% If your work has an appendix, this is the place to put it.
% \appendix

\end{document}